\documentclass[11pt]{article}
\pdfoutput=1
\usepackage{jheppub}

\usepackage{amssymb}
\usepackage{amsmath}
\usepackage{slashed}
\usepackage{graphicx}
\usepackage{longtable}
\usepackage{verbatim}
\usepackage{amsfonts}
\usepackage{epstopdf}
\usepackage{xcolor}
\usepackage{units}
\usepackage{flip-acronyms}
\usepackage{tikz}
\usepackage{tikzfeynman}
\usepackage{caption}
\usepackage{subcaption}

\newcommand{\be}{\begin{eqnarray}}
\newcommand{\ee}{\end{eqnarray}}
\newcommand{\bea}{\begin{eqnarray}}
\newcommand{\eea}{\end{eqnarray}}
\newcommand{\beq}{\begin{eqnarray}}
\newcommand{\eeq}{\end{eqnarray}}
\def\beqa{\begin{eqnarray}}
\def\eeqa{\end{eqnarray}}

\def\lsim{\mathrel{\rlap{\lower4pt\hbox{\hskip1pt$\sim$}}
    \raise1pt\hbox{$<$}}}         %less than or approx. symbol
\def\gsim{\mathrel{\rlap{\lower4pt\hbox{\hskip1pt$\sim$}}
    \raise1pt\hbox{$>$}}}         %greater than or approx. symbol

\def\R{\mathcal{R}}

\def\x{(\R,\vec r,t)}

\newcommand{\dd}{\text{d}}

\newcommand{\msq}{m^{\prime 2}}
\newcommand{\Ms}{M_\odot}
\newcommand{\vst}{v_\star}
\newcommand{\omt}{\omega_T}
\newcommand{\oml}{\omega_L}
\newcommand{\op}{\omega_p}

\newcommand{\lb}{\left(}
\newcommand{\rb}{\right)}
\newcommand{\dmpair}{\text{DM}+\overline{\text{DM}}}

\newcommand{\pol}{\text{pol}}

\begin{document}

%\vspace*{-10mm}

\title{Light(ly)-coupled Dark Matter in the keV Range: Freeze-In and Constraints}
\author{Jae Hyeok Chang${}^a$, Rouven Essig${}^a$, and Annika Reinert${}^b$}
\affiliation{${}^a$ C.~N.~Yang Institute for Theoretical Physics, Stony Brook, NY, USA}
\affiliation{${}^b$ Bethe Center for Theoretical Physics and Physikalisches Institut der 
Universit\"at Bonn, Nu{\ss}allee 12, 
 53115 Bonn, Germany}
\date{\today}

%\vspace*{1cm}

\emailAdd{jaehyeok.chang@stonybrook.edu}
\emailAdd{rouven.essig@stonybrook.edu}
\emailAdd{areinert@th.physik.uni-bonn.de}

\abstract{
Dark matter produced from thermal freeze-out is typically restricted to have masses above roughly 1 MeV. However, if the couplings are small, the freeze-in mechanism allows for production of dark matter down to keV masses. We consider dark matter coupled to a dark photon that mixes with the photon and dark matter coupled to photons through an electric or magnetic dipole moment.  We discuss contributions to the freeze-in production of such dark matter particles from standard model fermion-antifermion annihilation and plasmon decay. We also derive constraints on such dark matter from the cooling of red giant stars and horizontal branch stars, carefully evaluating the thermal processes as well as the bremsstrahlung process that dominates for masses above the plasma frequency.  We find that the parameters needed to obtain the observed relic abundance from freeze-in are excluded below a few tens of keV, depending on the value of the dark gauge coupling constant for the dark photon portal model, and below a few keV, depending on the reheating temperature for dark matter with an electric or magnetic dipole moment.  While laboratory probes are unlikely to probe these freeze-in scenarios in general, we show that for dark matter with an electric or magnetic dipole moment and for dark matter masses above the reheating temperature, the couplings needed for freeze-in to produce the observed relic abundance can be probed partially by upcoming direct-detection experiments.  
}

\preprint{YITP-SB-19-40, BONN-TH-2019-06}

\maketitle
%\tableofcontents

%%%%%%%%%%%%%%%%%%%%
\section{Introduction}

Dark matter with mass in the keV to GeV range has been receiving increased attention over the last few years.  
Numerous mechanisms exist for how such dark matter could have been produced in the early Universe. 
The mechanism of thermal freeze-out, which is perhaps the best studied mechanism, typically produces dark matter consistent with observations only between about $\sim$1~MeV to $\sim$100~TeV, being bounded below by bounds from Big Bang Nucleosynthesis (BBN)~\cite{Serpico_2004,Boehm:2013jpa,Green:2017ybv} and above by unitarity of the annihilation cross section~\cite{Griest:1989wd}.  
However, besides producing dark matter below the GeV scale by traditional thermal freeze-out~\cite{Boehm:2003hm}, various other related and non-thermal production mechanisms exist, see e.g.~\cite{Feng:2008ya,Hooper:2008im,Kaplan:2009ag,McDonald:2001vt,Hall:2009bx,Essig:2010ye,Essig:2011nj,Chu:2011be,Falkowski:2011xh,Hochberg:2014dra,Kuflik:2015isi,Izaguirre:2015yja,DAgnolo:2015ujb,Pappadopulo:2016pkp,Farina:2016llk,DAgnolo:2017dbv,DAgnolo:2018wcn,DAgnolo:2019zkf,Hambye:2019dwd,Heeba:2019jho,Battaglieri:2017aum,Evans:2019vxr,Koren:2019iuv}. 

In this paper, we consider several models for dark matter down to keV masses for which the relic abundance can be produced from freeze-in~\cite{McDonald:2001vt,Hall:2009bx}.  The couplings between the dark matter and Standard Model (SM) particles needed to obtain the observed relic abundance are typically very small, which naturally allows these models to avoid the BBN bound, since the dark matter particles will not be in chemical equilibrium with SM particles.  We show that nevertheless, at least for sufficiently low masses, these models can be probed by constraints from the cooling of various stellar objects. 

The models we consider are a dark matter particle coupled to a dark photon that mixes with the photon~\cite{Holdom:1985ag,Galison:1983pa,ArkaniHamed:2008qn,Pospelov:2008jd}, dark matter with an electric dipole moment (EDM), and dark matter with a magnetic dipole moment (MDM)~\cite{Bagnasco:1993st,Pospelov:2000bq,Sigurdson:2004zp,Banks:2010eh,Graham:2012su,Chu:2018qrm}. These models can naturally have small interactions with electrically charged SM particles through a small kinetic mixing parameter (for the dark photon portal) or through a higher dimension operator (for the EDM/MDM models). In these models, if the couplings are sufficiently small, the dark matter particles are never in thermal equilibrium with the SM particles, but are produced gradually from the SM thermal bath over time to produce the correct relic abundance. This is called the freeze-in mechanism. Depending on the type of interaction, the production may be dominant at low temperatures (IR freeze-in) or also occur at approved high temperatures (UV freeze-in)~\cite{Elahi:2014fsa}. The dark photon model has IR freeze-in, in which case the results do not depend on the reheating temperature, while the models with an electric or magnetic dipole moment have UV freeze-in, where the reheating temperature matters. We calculate the freeze-in parameters for these models, including the contributions from the plasmon decay~\cite{Braaten:1993jw,Redondo:2008ec,Dvorkin:2019zdi}.

The couplings needed for freeze-in are typically so small that these models cannot be constrained from laboratory experiments.  However, constraints from stellar objects, such as red giant stars (RG) and horizontal branch stars (HB), can probe these small couplings~\cite{Raffelt:1996wa,Redondo:2008aa,Redondo:2008ec,An:2013yfc,Vogel:2013raa,Redondo:2013lna,Chang:2016ntp,Knapen:2017xzo,Chang:2018rso,Chu:2019rok}. In stellar objects, SM particles can collide and produce the hypothetical dark sector particles. These dark sector particles can then carry away energy and change the evolution histories of the stellar objects. Since the observed stellar properties are consistent with predictions from the standard stellar models, we can constrain the couplings of the dark sector to the SM particles.  
Very roughly, since the temperature of the RG and HB stars reach about $10^8$~K, dark sector particle masses up to about 10~keV can be probed (more precisely, we will see that both the temperature and the plasma frequency inside the stars set the maximum mass that can be probed). 

The remainder of the paper is organized as follows.  In Sec.~\ref{sec:models}, we describe the salient features of the models considered in this paper and some basic constraints on them.  In Sec.~\ref{sec:FI}, we describe the freeze-in production in some detail.  
Sec.~\ref{sec:stellarconstraints} discusses the constraints from the RG stars and HB stars.  Sec.~\ref{sec:lab} briefly describes the prospects for probing these models in the laboratory.  We present our conclusions in Sec~\ref{sec:conclusions}. Three appendices provide additional details of our calculations.  

\section{Light Dark Matter Models Interacting or Mixed with Photons}\label{sec:models}
In this work we focus on dark matter interacting with photons, either via kinetic mixing through a heavy dark photon or directly due to an electric or magnetic dipole moment. 

For the dark photon ($A'$) portal, we will consider the dark matter candidate to be a fermion ($\chi$) or a complex scalar ($\phi$). The dark photon is the gauge boson of an additional broken U(1) gauge group, and it is kinetically mixed with the photon~\cite{Holdom:1985ag}. 
The Lagrangian is 
\begin{align}
\label{eq:lagrangian}
\mathcal{L}_{A'}=&-\frac{1}{4}F^\prime_{\mu\nu}F^{\prime\mu\nu}-\frac{\epsilon}{2\cos\theta_W}F^\prime_{\mu\nu}B^{\mu\nu}-\frac{1}{2}m^{\prime 2}A_\mu^\prime A^{\prime\mu}\\
&+
\begin{cases}
\bar{\chi}\left(i\gamma^\mu\partial_\mu +g_D\gamma^\mu A^\prime_\mu -m_\chi\right)\chi,\quad & \text{(Dirac fermion)}\\
\partial^\mu\phi\partial_\mu\phi^*-ig_DA^{\prime\mu}(\phi(\partial_\mu \phi^*)-(\partial_\mu\phi)\phi^*) +g_D^2A^{\prime 2}_\mu|\phi|^2 -m_\phi^2|\phi|^2\,,\quad & \text{(complex scalar)}\nonumber
\end{cases}
\end{align}
where $\epsilon$ is the kinetic mixing parameter, $\theta_W$ the weak mixing angle, $g_D=\sqrt{4 \pi \alpha_D}$ is the ``dark'' gauge coupling, and $B_{\mu\nu}$ and $F'_{\mu\nu}$ are the field strength tensors of the hypercharge gauge boson and the dark photon, respectively. If the dark photon is massless, $\chi$ or $\phi$ is a millicharged particle, for which the stellar cooling constraints have already been discussed in the literature~\cite{Davidson:2000hf,Vogel:2013raa,Chang:2018rso}.  Moreover, even if the dark photon is massive but ultralight $m'\ll m_\chi$, the constraints are similar to the millicharged case~\cite{Essig:2015cda,Hochberg:2015fth}. 
We therefore focus here on the ``heavy'' dark photon case, where $m'\sim \mathcal{O}(m_\chi)$. 
We mainly focus on the case with $m'>2m_\chi$ or $m'>2m_\phi$, in which case the dark matter will consist of $\chi$- or $\phi$-particles, and  we will only briefly comment on the case with $m'<2m_\chi$ or $m'<2m_\phi$, in which case the dark matter can mostly consist of dark photons. 

A model where the dark matter has an electric dipole moment ($d_\chi$) or a magnetic dipole moment ($\mu_\chi$) is described by the following term in the Lagrangian
\begin{align}\label{eq:Ledm}
\mathcal{L}_\text{EDM}&=-\frac{i}{2}d_\chi\bar{\chi}\sigma_{\mu\nu}\gamma^5\chi F^{\mu\nu}\ ,\\
\mathcal{L}_\text{MDM}&=-\frac{1}{2}\mu_\chi\bar{\chi}\sigma_{\mu\nu}\chi F^{\mu\nu}\ ,
\end{align}
respectively.
The $d_\chi$ and $\mu_\chi$ have mass dimension $-1$. These effective operators must come from an underlying theory at a larger scale. As an example, the dipole moment can be induced by heavy charged particles (a fermion and a scalar) that couple the dark matter to the SM through a loop~\cite{Graham:2012su}. In such a scenario, for charged particles of mass $M$, the electric dipole moment would be given by 
\begin{equation}
d_\chi \sim \frac{e g^2}{8\pi^2 M}\ ,
\end{equation}
where $e$ is the electron charge and $g$ is the coupling between the heavy charged particles and $\chi$.  A similar equation holds for $\mu_\chi$.  The mass of the heavy charged particles $M$ could be as light as $\sim$100~GeV~\cite{Tanabashi:2018oca} (possibly even slightly smaller~\cite{Egana-Ugrinovic:2018roi}) to avoid collider bounds, which can be combined with a limit of $g^2<4\pi$ requiring perturbativity to give an upper bound of $d_\chi\lesssim0.5~\text{TeV}^{-1}$.  Note that this limit is stronger than the LEP limit on dark matter particles with electric or magnetic dipole moment directly, which is about $d_\chi\lesssim 4~\text{TeV}^{-1}$~\cite{Fortin:2011hv} for $m_\chi<50$~GeV. 
However, in this simple model of an additional scalar and fermion generating the dipole moment, the dark matter mass also receives loop corrections of roughly~\cite{Graham:2012su}
\begin{align}
\delta m_\chi\sim\frac{M^2}{2e}d_\chi\ .
\end{align}
Since the dark matter mass cannot (trivially) be smaller than its mass correction, we find an upper limit for the electric dipole moment of 
\begin{align}\label{eq:dipole-upper-bound}
d_\chi \lesssim 10^{-5}\left( \frac{m_\chi}{1~\text{MeV}} \right) \left (\frac{1~\text{TeV}}{M}\right )^2 \ \text{TeV$^{-1}$}\ ,
\end{align}
with a similar equation for the magnetic dipole moment.  
Of course, one could imagine different UV completions of the dark matter models with a dipole moment that do not have the same strong upper bound on the dipole moment.

\subsection*{General bounds on keV-to-GeV mass dark matter}

Our main focus in this paper is on deriving the stellar constraints and freeze-in production of dark matter in the keV to GeV mass range.  
However, in the remainder of this section we briefly review other bounds on dark matter in or near this mass range. 

If the dark matter is in chemical equilibrium with the SM bath in the early Universe, dark matter masses below $\sim$9.4~MeV (for a Dirac fermion) or $\sim$6.5~MeV (for a complex scalar)~\cite{Boehm:2013jpa,Green:2017ybv,Sabti:2019mhn} are in tension with cosmological observables. The reason is that in the cosmological standard model, BBN started at a temperature of about $1$~MeV, and the predictions are well confirmed by the measured abundance of light elements; extra relativistic degrees of freedom, $N_{\rm eff}$, during this evolutionary stage could affect the expansion of the Universe and thus change the temperature at which BBN begins, which would alter the predicted values for the abundance of light elements. 

If the particles were never in chemical equilibrium with the SM (which is a necessary condition for freeze-in production), the number density is much smaller than the equilibrium number density, and the contribution to $N_{\rm eff}$ is negligible. 
We will check this condition below when we compare the parameters needed for freeze-in production to the couplings that would keep the dark matter in chemical equilibrium with the SM bath. 

Another constraint on the dark matter mass comes from the existence of small-scale structure.  Below dark matter masses of 1~keV,  fermionic dark matter cannot account for all of the dark matter in dwarf galaxies~\cite{Tremaine:1979we,Boyarsky:2008ju} due to the Pauli principle.  Moreover, when the first structures form, the process can be disturbed if (a large component of) the dark matter (either fermionic or bosonic) is too warm. The fast streaming of the particles would then `wash out' the forming structures. For thermal relics this is the case if the mass is lighter than about 1~keV~\cite{Gorbunov:2011zz}. In general, the constraint depends on the momentum distribution of the dark sector, which is modified in the non-thermal case by the average momentum $\langle |\textbf{p}|\rangle$ compared to the thermal momentum $\langle |\textbf{p}|\rangle^\text{eq}$~\cite{Gorbunov:2011zz}
\begin{align}
m_{\chi/\phi}\geq \frac{\langle |\textbf{p}|\rangle}{\langle |\textbf{p}|\rangle^\text{eq}}\times 1\ \text{keV}\ .
\end{align}
While a detailed calculation of the resulting mass bound in our models is beyond the scope of this paper, we do not expect the result to differ significantly from 1~keV (see e.g.~\cite{Dvorkin:2019zdi}, which considered a model consisting of dark matter interacting with a very light dark photon and find only an $\mathcal{O}$(1) correction; see also~\cite{Kamada:2019kpe}). 

For the dark-photon portal, large values of $\alpha_D$ imply large self-interactions among the dark matter particles mediated by the dark photon.  For dark photons with a mass near the dark matter mass, the self-interaction limit on $\alpha_D$ from observations of the Bullet cluster, $\sigma_\text{SIDM}/m_\chi\lesssim 2$~cm$^2$/g~\cite{Robertson:2016xjh}, become very stringent for small masses.  
In particular, for $m'=3m_\chi$, we need $\alpha_D<0.5$ for dark matter masses below $\sim$20~MeV, while we need $\alpha_D\leq 10^{-6}$ for dark matter masses below $\sim$28~keV.  We will see below that the stellar constraints disfavor dark matter interacting with a dark photon to constitute a dominant component of dark matter from freeze-in production for dark matter masses below approximately 15~keV for $\alpha_D = 10^{-6}$.  Below, we will consider values for $\alpha_D$ ranging between $10^{-6}$ to 0.5.  
For dark matter interacting through a dipole moment, the self-interaction limits are not relevant, since the couplings to the mediator---the photon in this case---are very small.

We note that the bounds from $N_{\rm eff}$, structure formation, and self-interactions may be evaded if the dark matter candidates constitute only a sub-component of the observed dark matter density.  Laboratory bounds will be discussed in Sec.~\ref{sec:lab}.

\section{Production via Freeze-In}\label{sec:FI}
If the interaction between the dark sector and SM sector is sufficiently small, the dark sector was never in chemical equilibrium with the SM sector throughout the history of the Universe. Excluded from the thermal bath, the dark matter abundance today therefore cannot be set via the typical freeze-out mechanism. Still, as long as there is some small coupling between the SM and dark sector, SM particles in the thermal bath can annihilate to produce dark sector particles. This is called the freeze-in mechanism~\cite{McDonald:2001vt,Hall:2009bx}, and here we consider the freeze-in production of dark matter interacting with a heavy dark photon mediator, an electric dipole moment, or a magnetic dipole moment, as discussed in Sec.~\ref{sec:models}. 

In general, the number density, $n_a$, of a particle species $a$ produced within the thermal history of the Universe is derived from the Boltzmann equation
\be
\frac{dn_a}{dt}+3Hn_a=R(T)\ .
\ee
Here, $R(T)$ is the number of interactions per unit volume and per unit time in which the particle is produced,
\begin{equation}
\label{eq:RT}
R(T) = \int \dd \Pi_i \dd \Pi_f |\mathcal{M}_{i \rightarrow f}|^2 (2\pi)^4 \delta^{(4)} \left( \Sigma p_i - \Sigma p_f \right)\,,
\end{equation}
where
\begin{equation}\label{eq:dPi}
\dd \Pi_i=\displaystyle \prod_{i} \frac{g_i d^3p_i}{(2\pi)^3 2E_i} f_i, ~~\dd \Pi_f=\displaystyle \prod_{f} \frac{g_f d^3p_f}{(2\pi)^3 2E_f} (1 \pm f_f)\, ,
\end{equation}
$p$ and $E$ are the momentum and energy, respectively, subscripts $i$ and $f$ correspond to initial and final particles, $g$ is the spin degeneracy, $f$ is the distribution function, $+$ and $-$ correspond to bosons and fermions, respectively, and the final state includes the particle $a$. $R(T)$ for $2 \rightarrow 2$ processes is conventionally written as $n_i^2 \langle\sigma v\rangle$, and for $1 \rightarrow 2$ processes it is written as $n_i \langle \Gamma \rangle$.

The yield $Y$ from freeze-in is found by integrating \cite{Dutra:2018gmv}
\begin{align}\label{eq:dYdT}
\frac{\dd Y}{\dd T}=-2\frac{M_\text{Pl}}{(2\pi)^2}\left(\frac{45}{\pi}\right)^{3/2}\frac{\tilde{g}}{g^*_s \sqrt{g^*}}\frac{R(T)}{T^6}\ .
\end{align}
Here, $M_\text{Pl}\simeq1.22\times 10^{22}$~MeV is the Planck mass, $g^*(T)$ is the effective number of relativistic degrees of freedom at temperature $T$, $g^*_s(T)$ the entropic relativistic degrees of freedom, and $\tilde{g}(T)=\left(1+\frac{T}{3}\frac{\dd \ln(g^*_s)}{\dd T}\right)$ (see e.g.~\cite{Husdal:2016haj}). The first factor of two in Eq.~\eqref{eq:dYdT} accounts for the production of both particles and antiparticles.

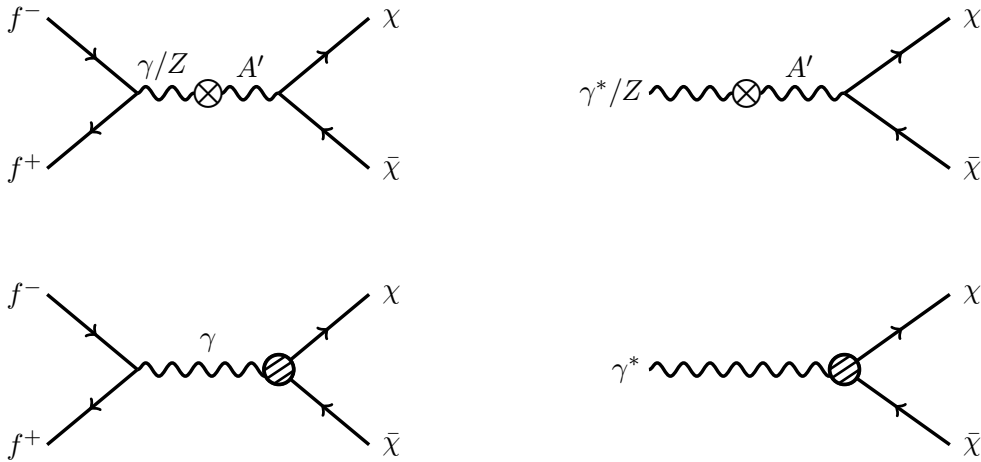
\begin{figure}
\centering
\begin{tikzpicture}[line width=1.3 pt, scale=2]
	\draw[fermion] (0,0.5)--(.6,0);
	\draw[fermionbar] (0,-.5)--(.6,0);
	\draw[vector] (.6,0)--(0.98,0);
	\draw[vector] (1.16,0)--(1.54,0);	
	\draw[fermion] (1.54,0)--(2.14,.5);
	\draw[fermionbar] (1.54,0)--(2.14,-.5);

	\node at (1.07, 0) {$\bigotimes$};
	\node at (-.15, -.5) {$f^+$};
	\node at (-.15,.5) {$f^{-}$};
	\node at (0.77,0.18) {$\gamma/Z$};
	\node at (1.35,0.18) {$A'$};
	\node at (2.29,.5) {$\chi$};
	\node at (2.29,-.5) {$\bar\chi$};

	\draw[vector] (4,0)--(4.56,0);
	\draw[vector] (4.74,0)--(5.3,0);
	\draw[fermionbar] (5.3,0)--(6,-.5);
	\draw[fermion] (5.3,0)--(6,.5);
	
	\node at (4.65, 0) {$\bigotimes$};
%	\node at (4.3,.18) {$\gamma$};
	\node at (3.75,0) {$\gamma^*/Z$};
	\node at (5,0.18) {$A'$};
	\node at (6.15,.5) {$\chi$};
	\node at (6.15,.-.5) {$\bar\chi$};
\end{tikzpicture}

\vspace{1 cm}

\begin{tikzpicture}[line width=1.3 pt, scale=2]
	\draw[fermion] (0,0.5)--(.6,0);
	\draw[fermionbar] (0,-.5)--(.6,0);
	\draw[vector] (.6,0)--(1.54,0);
	\draw[fermion] (1.54,0)--(2.14,.5);
	\draw[fermionbar] (1.54,0)--(2.14,-.5);

	\draw[fill=black] (1.54,0) circle (.1cm);
	\draw[fill=white] (1.54,0) circle (.098cm);
	\begin{scope}
    	\clip (1.54,0) circle (.1cm);
    	\foreach \x in {-.9,-.8,...,2.}
			\draw[line width=1 pt] (\x,-.1) -- (\x+.3,.1);
  	\end{scope}

	\node at (-.15, -.5) {$f^+$};
	\node at (-.15,.5) {$f^{-}$};
	\node at (1.07,0.18) {$\gamma$};
	\node at (2.29,.5) {$\chi$};
	\node at (2.29,-.5) {$\bar\chi$};

	\draw[vector] (4,0)--(5.3,0);
	\draw[fermionbar] (5.3,0)--(6,-.5);
	\draw[fermion] (5.3,0)--(6,.5);
	
	\draw[fill=black] (5.3,0) circle (.1cm);
	\draw[fill=white] (5.3,0) circle (.098cm);
	\begin{scope}
    	\clip (5.3,0) circle (.1cm);
    	\foreach \x in {3.85,3.95,...,6.15}
			\draw[line width=1 pt] (\x,-.1) -- (\x+.3,.1);
  	\end{scope}	
	
	\node at (3.85,0) {$\gamma^*$};
	\node at (6.15,.5) {$\chi$};
	\node at (6.15,.-.5) {$\bar\chi$};
\end{tikzpicture}
 
	\caption{The dominant processes relevant for the production of dark matter interacting with a dark photon (upper diagrams) and dipole moments (lower diagrams) in the early Universe: pair annihilation (left) and plasmon/$Z$-boson decay (right).}
	\label{fig:FeynFI}
\end{figure}

For the production of light dark matter coupled to the SM via the dark photon, there are several processes: pair 
annihilation of the SM particles $f$ with $\bar{f}$, plasmon decay, and $Z$-boson decay, 
\begin{align}\label{eq:ratedp}
R(T)_\text{dark photon}= \sum_f n_f^2\langle\sigma v\rangle_{f\bar{f}\rightarrow \dmpair}+n_{\gamma^\star}\langle\Gamma\rangle_{\gamma^\star\rightarrow\dmpair} + n_{Z}\langle\Gamma\rangle_{Z\rightarrow\dmpair}\ .
\end{align}
The contribution from $Z$-boson decay is important for $m_\chi \gtrsim 1~\textrm{GeV}$. For $m_\chi \lesssim 1~\textrm{GeV}$, which is our focus in this paper, the $Z$-boson decay contributes  $\mathcal{O}(10\%)$.  
For the production of light dark matter coupled to the SM via an electric or magnetic dipole moment, there are pair 
annihilation of the SM particles $f$ with $\bar{f}$ and plasmon decay processes that contribute to the production rate  
\begin{align}\label{eq:rateedmmdm}
R(T)_\text{EDM/MDM}= \sum_f n_f^2\langle\sigma v\rangle_{f\bar{f}\rightarrow \dmpair}+n_{\gamma^\star}\langle\Gamma\rangle_{\gamma^\star\rightarrow\dmpair}\ .
\end{align}
We show relevant processes for a dark matter fermion coupled to a dark photon or with an electric or magnetic dipole moment in Fig.~\ref{fig:FeynFI} (the case of a scalar dark matter particle coupled to a dark photon is similar).  

We can estimate from dimensional analysis the temperature at which dark matter production via freeze-in is important. Since $R(T)$ has mass dimension 4, $R(T) \sim T^4$ for dimension-4 operators like the dark photon kinetic mixing term, and $R(T) \sim \frac{T^6}{\Lambda^2}$ for dimension-5 operators like the electric or magnetic dipole moment interactions.  For the dark photon case, this then implies that $\frac{dY}{dT} \sim T^{-2}$, and freeze-in production is dominant at low temperature and is not sensitive to the reheating temperature. It is thus said to be infrared-dominated. In contrast, for the dipole moment case, $\frac{dY}{dT} \sim T^{0}$, so processes at all temperatures are relevant, up to the reheating scale, which therefore determines the relic abundance. 

For the infrared dominated dark photon case, the precise process that dominates the freeze-in production depends on the dark matter mass. If the dark matter is heavier than the electron mass, pair annihilation dominates over plasmon decays (both longitudinal and transverse plasmons).  Transverse plasmon decays occur at higher temperatures than the annihilation process, as the plasma frequency has to fulfill $2 m_\chi \lesssim \omega_p\approx T/10$.  The plasmon production process thus becomes inefficient for $T\lesssim 20 m_\chi$, while the pair annihilation process is still very efficient.  For dark matter masses above the electron mass, the dominant contributions are expected from those SM particles that are lighter than the dark matter and freeze out only after the freeze-in production has been completed.  For example, for $m_e \lesssim m_\chi \lesssim m_\mu$,  the dark matter freeze-in production is dominated by electron-positron pair annihilation. 
On the other hand, for dark matter masses below the electron mass, the pair annihilation process quickly becomes inefficient, and the decay of (transverse) plasmons yields sizable contributions to the number density.  
The longitudinal plasmon modes are always suppressed compared to the transverse plasmon modes for infrared dominated production due to lack of available phase space. 

For the production of dark matter with an electric or magnetic dipole moment, which is determined by the reheating scale, plasmon decays are always sub-dominant. This is a general conclusion for UV-freeze-in through a high-dimensional operator. 

We now consider the two production processes---pair annihilation and plasmon decay---in more detail below. 
For the subdominant $Z$-boson decay contribution, we provide detailed formulae in Appendix~\ref{app:Z-boson}. 
Matching the dark matter density to the observed relic abundance today, $\Omega_\text{DM} h^2\simeq0.11$~\cite{PhysRevD.98.030001}, will allow us to find the value of the kinetic mixing factor, $\epsilon$, or the electric or magnetic dipole moment, $d_\chi$ or $\mu_\chi$, that gives the right relic abundance. 

\subsection{Pair Annihilation}
For dark matter masses above $\sim$1~MeV, the freeze-in production is dominated by contributions from annihilation into the dark sector. The rate to produce the dark matter particle $f$ in a thermal bath of temperature $T$ can then be derived from~\cite{Gondolo:1990dk,Dutra:2018gmv}
\begin{align}\label{eq:rateElScatt}
n_f^2\langle\sigma v\rangle_\text{prod}=\frac{g_i g_j T}{32(2\pi)^6}\int\dd s\sqrt{s}K_1\left(\frac{\sqrt{s}}{T}\right)\sqrt{1-\frac{4m_{\chi/\phi}^2}{s}}\sqrt{1-\frac{4m_f^2}{s}}\int\dd\Omega\left|\mathcal{M}\right|^2\ ,
\end{align}
where $m_f$ is the mass of particle $f$, $K_n(x)$ is the modified Bessel function of the second kind, and $s$ is the usual Mandelstam variable; $g_i=g_j=2$ account for the degrees of freedom of the incoming fermions.
Here $|\mathcal{M}^2|$ is averaged over initial state spins and summed over final state spins. 
We can now consider several cases: 
\begin{itemize}
\item
If the dark matter is a fermion interacting with a dark photon, the integrated matrix element of the annihilation of SM fermions $f+\bar{f}$ with charge $q_f$ in units of $e$ and mass $m_f$ is given by~\cite{Dutra:2018gmv}
\begin{align}\label{eq:intMatrixScattFermion}
\int\dd\Omega\left|\mathcal{M}\right|^2=\frac{16 \pi}{3}(\epsilon e q_f g_D)^2\left(1+\frac{2m_f^2}{s}\right)\left(1+\frac{2m_\chi^2}{s}\right)\frac{s^2}{(s-m^{\prime 2})^2+m^{\prime 2}\Gamma_{A^\prime}^2}\,,
\end{align}
with the total width of the dark photon given by~\cite{Dutra:2018gmv}\footnote{We correct the result given in~\cite{Dutra:2018gmv} by a factor of $1/3$ to account for the average over the three polarization modes.} 
\begin{align}\label{eq:widthFermion}
\Gamma_{A^\prime}=\frac{m^{\prime}}{12\pi}\left[g_D^2\lb 1+\frac{2m_\chi^2}{m^{\prime 2}}\rb\sqrt{1-\frac{4 m_\chi^2}{m^{\prime 2}}} + \sum_f (\epsilon e q_{f})^2\lb 1+\frac{2m_{f}^2}{m^{\prime 2}}\rb\sqrt{1-\frac{4m_{f}^2}{m^{\prime 2}}}\right].
\end{align}
These formulae do not include $A'$-$Z$-mixing, but we show the full formulae that include this mixing in Appendix~\ref{app:Z-boson}. 
In the second term of Eq.~\eqref{eq:widthFermion}, the sum is over all fermions that are lighter than the dark photon. If $m^{\prime}<2 m_e$, only the first term contributes. 
In our calculations, we drop the second term, which is a reasonable approximation as long as $\alpha_D=g_D^2/4\pi \gg \alpha \epsilon^2$. 
For Eq.~\eqref{eq:intMatrixScattFermion}, the major contribution comes from electron-positron annihilations, since the freeze-in is dominated by the lowest temperatures. 

\item
If the dark matter is a boson interacting with a dark photon, the integrated matrix element is
\begin{align}\label{eq:matrixBoson}
\int\dd\Omega\left|\mathcal{M}\right|^2=\frac{4\pi}{3}(\epsilon e q_f g_D)^2\left(1+\frac{2m_f^2}{s}\right)\left(1-\frac{4 m_\phi^2}{s}\right)\frac{s^2}{(s-\msq)^2+\msq\Gamma_{A^\prime}^2}\ ,
\end{align}
and the total width of the dark photon is
\begin{align}\label{eq:width}
\Gamma_{A^\prime}=\frac{m^\prime}{12\pi}\left[\frac{g_D^2}{4}\left(1-\frac{4m_\phi^2}{\msq}\right)^{3/2} + \sum_f(\epsilon e q_{f})^2\left(1+\frac{2m_{f}^2}{\msq}\right)\sqrt{1-\frac{4m_{f}^2}{\msq}}\right]\ .
\end{align}
Again, the second term accounts for the dark photon decays to SM particles, and the first term accounts for the decay to dark matter particles. A factor of $1/3$ appears from averaging over the dark photon polarizations.  We again drop the second term in our calculations. 

\item
For the case of dark matter with an electric or magnetic dipole moment, the integrated matrix elements are
\begin{align}
\int\dd\Omega\left|\mathcal{M}^\text{EDM}\right|^2 & = d_\chi^2\frac{32\pi^2\alpha}{3}\frac{(s+2m_f^2)(s-4m_\chi^2)}{s}\,\\
\int\dd\Omega\left|\mathcal{M}^\text{MDM}\right|^2 & =  \mu_\chi^2\frac{32\pi^2\alpha}{3}\frac{(s+2m_f^2)(s+8m_\chi^2)}{s}\ ,
\end{align}
respectively. 
\end{itemize}
With these expressions, Eq.~\eqref{eq:rateElScatt} can be used to derive the number density averaged interaction rate that enters Eq.~\eqref{eq:dYdT} to yield the relic abundance.

\subsection{Plasmon decay}\label{sec:plasmonMatrixEl}
Another important production process for dark matter comes from the decay of plasmons in the thermal plasma of the early Universe. 
In a thermal plasma, the interaction of the photon with charged particles, most dominantly electrons, leads to an effective mass for the photon, which depends on the electron density and temperature of the thermal bath (see Appendix~\ref{app:plasmaProp}).  
At finite temperature, the photon propagator gets renormalized. The additional term acts like a self-energy of the photon, making it effectively massive. These quasi-massive states are called plasmons. The pole of the photon propagator determines the dispersion relations, which are then modified in comparison to the vacuum case. 
The properties of plasmons differ significantly from photons propagating in vacuum: they move slower than the speed of light and there is a longitudinal mode in addition to the transverse modes. We will refer to the different modes as `longitudinal' or `transverse' plasmons. 

The plasmon production and decay includes all electromagnetic processes where on-shell photons are produced. 
One thus has to be careful to avoid double counting diagrams that contribute with an on-shell photon. However, here there is no such danger for the annihilation process; the intermediate photon cannot be on-shell, which can be seen by cutting the diagram (see Fig.~\ref{fig:FeynFI}) in the middle. If the photon were on-shell, the inverse process of the right-hand-side would correspond to a plasmon that decays to an electron-positron pair. 
This process is kinematically not allowed for plasmons, since the charged SM particles receive corrections to their mass and are too heavy. 
Thus, in the early Universe, the decay of plasmons and the annihilation of SM particles are two distinct processes that can be treated separately. 

The decay of plasmons is an important production mechanisms for neutrinos in stellar objects~\cite{Braaten:1993jw}. Similarly, the plasmons can decay into particles from a dark sector. This process is relevant in the early Universe as well as in stellar objects. In the following, we follow the notation and conventions of~\cite{Braaten:1993jw}. 

We now want to derive the production rates of dark matter from plasmon decay.
In general, the thermally averaged rate is given by
\begin{equation}\label{eq:ratePlasmon}
n_{\gamma^\star} \langle\Gamma\rangle_{\gamma^\star\rightarrow \dmpair} = \sum_\pol\int \frac{\dd^3k}{(2\pi)^3} g_\pol f(\omega_\pol) \Gamma^\pol_{\gamma^\star\rightarrow \dmpair}
\end{equation}
\begin{eqnarray}
\Gamma^\pol_{\gamma^\star\rightarrow \dmpair} &=& \frac{1}{2\omega_\pol} \int \frac{\dd^3 p_\text{DM}}{(2\pi)^3 2E_\text{DM}}\frac{\dd^3 p_{\overline{\text{DM}}}}{(2\pi)^3 2E_{\overline{\text{DM}}}}(2\pi)^4\delta^{(4)}\left(K-P_\text{DM}-P_{\overline{\text{DM}}}\right)|\mathcal{M}_\pol|^2_{\gamma^\star\rightarrow\dmpair}\nonumber \label{eq:PlasmonGamma}\\
\\
&=& \frac{Z_\pol}{16 \pi} \sqrt{1-\frac{4m_\text{DM}^2}{\omega_\pol^2-k^2}} \frac{f_\text{DM}(\omega_\pol^2-k^2)}{\omega_\pol} \label{eq:plasmondecayrate}
\end{eqnarray}
Here, $K=(\omega,\vec{k})$ is the four momentum of the plasmon and $P_\text{DM}, P_{\overline{\text{DM}}}$ are the four momenta of the outgoing dark matter particles. The sum over the polarizations `$\pol$' includes one longitudinal ($L$) and two transverse ($T$) modes, thus $g_L=1$ and $g_T=2$.  The distribution function of the plasmons is a Bose-Einstein distribution $f(\omega)=1/(\exp(\omega/T)-1)$, and $Z_\pol$ (i.e., $Z_L$ and $Z_T$) is given in Appendix~\ref{app:plasmaProp}. To get Eq.~\eqref{eq:PlasmonGamma}, we follow the calculations in~\cite{Chu:2019rok} for integrating over the outgoing dark matter phase space. The corresponding 
$f_{\text{DM}}(s)$, with $s=\omega_\pol^2-k^2$, for each model that we consider is
\begin{eqnarray}
f_\text{DM}^{A'}(s) &=& \frac{16 \pi}{3} \epsilon^2 \alpha_D \frac{s^2 (s+2 m_\chi^2)}{(s-m'^{2})^2+m'^{2} \Gamma_{A'}^2}\label{eq:fDMDP}\,,\\
f_\text{DM}^{\text{EDM}}(s) &=& \frac{2}{3} d_\chi^2 s(s-4m_\chi)^2\label{eq:fDMEDM}\,,\\
f_\text{DM}^{\text{MDM}}(s) &=& \frac{2}{3} \mu_\chi^2 s (s+8m_\chi^2)\,.\label{eq:fDMMDM}
\end{eqnarray}

%In the following subsections, we give the expressions for the matrix element in Eq.~\eqref{eq:ratePlasmon} for the models considered in this paper. 
The two processes indicated by the expressions in Eq.~\eqref{eq:rateElScatt} and Eq.~\eqref{eq:ratePlasmon} together then give the total production rate of the freeze-in process that appears in Eq.~\eqref{eq:dYdT}, which gives the final dark matter yield. It turns out that for  dark matter interacting with a dark photon mediator, the decay of the longitudinal mode contributes only at the percent level compared to the contribution from the transverse mode. However, for the dipole moment models, the contribution from the longitudinal mode can dominate. 

\subsection{Results}

\subsubsection{Dark photon + fermion dark matter} 
\begin{figure}
\center
	\includegraphics[width=0.495\textwidth]{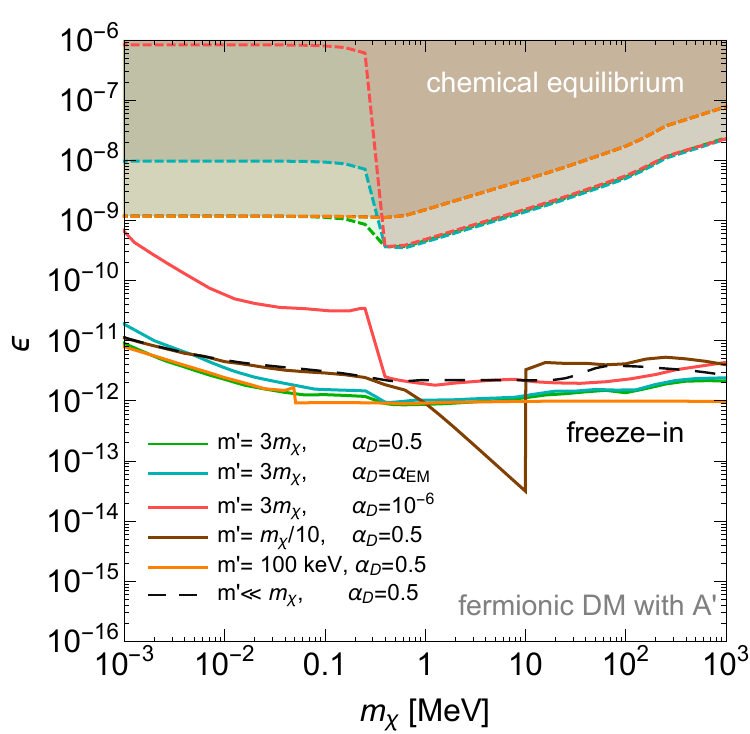}
	\includegraphics[width=0.495\textwidth]{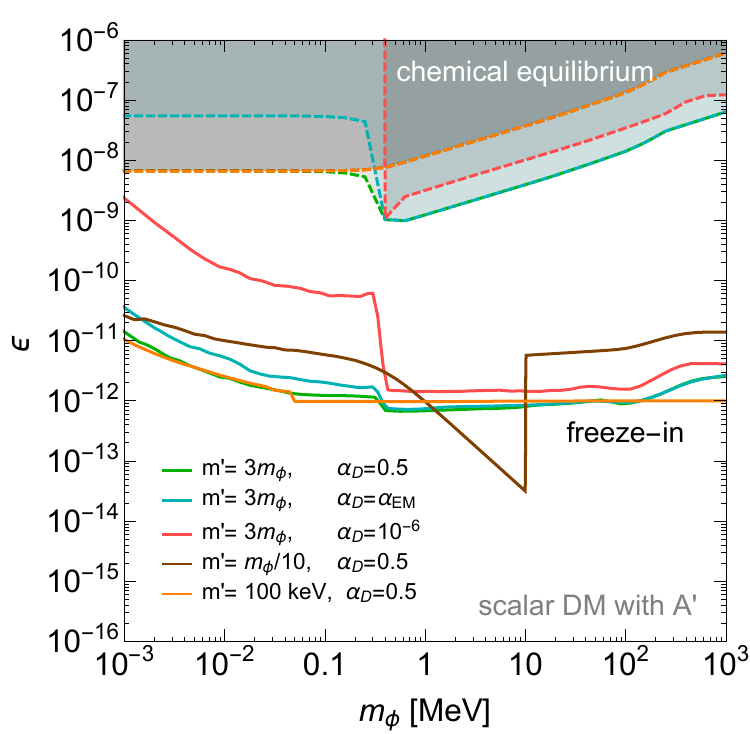}
	\caption{
Solid lines show the values of $\epsilon$ for which the correct dark matter relic abundance is obtained in a model with fermionic (left) and scalar (right) dark matter from freeze-in through a dark photon.  We show various choices for the dark-matter-to-dark-photon mass ratio and 
dark-photon couplings $\alpha_D$. 
For $m'<2 m_{\chi/\phi}$ and $m'<2m_e$, the dark photon can make up the relic abundance, rather than the fermion $\chi$ or scalar $\phi$ (see text for details). 
For parameters above these ``freeze-in lines'', too much dark matter is produced in the early Universe. Above the dashed lines, the dark matter and SM sector are in chemical equilibrium.  Below the chemical equilibrium lines the model is safe from constraints on the number of relativistic degrees of freedom in the early Universe.}\label{fig:FIfermion}
\end{figure}

In Fig.~\ref{fig:FIfermion} (left), we show the values of $\epsilon$ needed to obtain the correct relic abundance from freeze-in for the dark photon portal for a fermionic dark matter candidate.  We consider six different scenarios. We see that the value of $\epsilon$ that yields the correct relic abundance through the freeze-in mechanism does not depend very strongly on the model parameters and mass ratios.

We first discuss the case for which the dark photon can decay to dark matter, and choose $m^\prime=3m_\chi$.  We show lines corresponding to three different values of $\alpha_D$, namely $\alpha_D=0.5$, $\alpha_D=\alpha_\text{EM}\simeq1/137$, and $\alpha_D=10^{-6}$.  The latter value satisfies the self-interaction  bound down to dark matter masses of $\sim$28~keV, below which production via freeze-in is constrained from stellar cooling, see Sec.~\ref{sec:stellarconstraints}. 

From the matrix element for the annihilation Eq.~\eqref{eq:intMatrixScattFermion} or the plasmon decay Eq.~\eqref{eq:fDMDP}, we see that far off the resonance, where the decay width is negligible, the production rate is proportional to $\alpha_D\epsilon^2$. In contrast, close to the resonance, the coupling $\alpha_D$, which appears also in the decay width, divides out. In the early Universe, a range of temperatures is scanned, so the production always gets large contributions from the resonance at some point. It is for this reason that the scaling of $\epsilon$ between the different lines is less than a factor of $1/\sqrt{\alpha_D}$. 

If the dark matter is heavier than electrons, it is mainly produced through the annihilation process, which rapidly becomes inefficient when the temperature drops below the mass of a dark matter pair.  The production from plasmon decays is subdominant, since the plasma frequency, which is the measure for the available phase space of the decay, is much lower than the temperature, $\omega_p\approx T/10$. So the plasmon process contributes less than the annihilation process, since the freeze-in production of dark-photon-mediated dark matter particles is infrared dominated and the plasmon decay process stops at larger temperatures than the annihilation process. 

In contrast, for dark matter masses below the electron mass, the dominant contribution comes from plasmon decays. Again, thanks to the infrared domination, the main production happens at late times and stops when the production of dark matter is kinematically forbidden. If this occurs after electrons freeze out, the annihilation process does not contribute anymore. However, plasmon decays can still occur until the plasma frequency $\omega_p\approx T/10$ falls below twice the dark matter mass. 

We also present two cases where the dark photon is (partly) lighter than twice the dark matter mass. In this case, the dark photons can make up the entire dark matter abundance for $m^\prime < 2m_e$.  This scenario is studied in~\cite{Redondo:2008ec}, and we find that it does not receive significant corrections from the presence of heavier particles in the dark sector. 
However, we briefly discuss the composition of the dark relics in these scenarios. 
The small kink in the orange line ($m'=100$~keV) in Fig.~\ref{fig:FIfermion} marks the transition when the dark fermions become heavier than the dark photons.  While the dark photons are efficiently produced in the early Universe for all dark fermion masses, they rapidly decay to the dark fermions for dark fermion masses below the kink.  On the other hand, for dark fermion masses above the kink, the dark photons are very long-lived and constitute the dark relics. In the same figure, the brown line shows the case where the dark matter is always lighter than the dark photon ($m^\prime=m_\chi/10$). At dark photon masses below $\sim$100~keV, only a small number of dark photons is produced directly, such that their initial relic abundance is small. Thus, most of the dark relics that are frozen-in are dark fermions, which may then annihilate into dark photons. However, above roughly $m^\prime\sim100$~keV and below $m^\prime=2m_e$, the direct freeze-in production of dark photons becomes sizable compared to dark fermions, and they can constitute the dark relics.  For  $m^\prime>2 m_e$, the dark photon is short-lived and decays rapidly into electron positron pairs. This causes the big kink in the brown line in Fig.~\ref{fig:FIfermion}. 

Irrespective of the component that is most dominantly produced by the freeze-in mechanism (dark fermions or dark photons), processes like dark photon annihilation into dark fermions or vice versa could allow for a change in the relative abundance of the two species after freeze-in.  This will only occur if the dark gauge coupling is large enough and the two species achieve chemical equilibrium in the dark sector.  This has important implications for direct and indirect dark matter searches.  This is reminiscent of the ``leak-in'' scenario discussed in~\cite{Evans:2019vxr,Krnjaic:2017tio}.  
We leave further investigation of this effect to future work. 

Finally, we show the values of $\epsilon$ needed for the ultralight dark photon mediator case~\cite{Essig:2011nj,Chu:2011be,Essig:2015cda,Dvorkin:2019zdi}.  We see that these values are similar to the other cases, especially (at low dark matter masses) the case with $m^\prime=m_\chi/10$. 

When $\epsilon$ is smaller than the values indicated by the solid curves in Fig.~\ref{fig:FIfermion}, the relic abundance of $\chi$-particles and/or dark photons is less than the observed dark matter relic abundance, so that they form a subdominant dark matter component. Above the freeze-in line, too much dark matter would have been produced through freeze-in, and the Universe would be overclosed. We also show the lines indicating the coupling at which the dark matter would attain chemical equilibrium with the SM bath. This happens if the interaction rate $\Gamma=n_\chi^\text{eq}\langle\sigma v\rangle$ for the annihilation process is at any time bigger than the Hubble rate $H=\frac{\pi\sqrt{g_\text{eff}}}{\sqrt{90}M_\text{Pl}}T^2$. 
For the interaction rate, the annihilation into electrons is the most relevant process, i.e., the one that gives the most stringent constraint, and the rate is given by Eq.~\eqref{eq:rateElScatt} divided by the equilibrium number density~\cite{Gondolo:1990dk}
\begin{align}
n^\text{eq}_{i}=\frac{g_{i}}{2\pi^2}m^2_{i} T K_2\left(\frac{m_{i}}{T}\right)\ .
\end{align}
Here, $g_{i}$ is the number of degrees of freedom of the particle species; for a Dirac fermion or a complex scalar, $g_i=4$.
Chemical equilibrium would of course spoil the freeze-in mechanism, and it is thus important to check that the freeze-in line does not get too close to the coupling that allows for chemical equilibration. 
In Fig.~\ref{fig:FIfermion} (left), we see that for the dark photon portal dark matter this requirement is fulfilled. 

\subsubsection{Dark photon + scalar dark matter}
The $\epsilon$ values needed to obtain the correct relic abundance from freeze-in for scalar dark matter coupled to a dark photon are shown with solid lines in Fig.~\ref{fig:FIfermion} (right).  We also show with dashed lines the $\epsilon$ values above which chemical equilibrium with the SM is reached.   We find that the results for scalar dark matter look very similar to the fermion dark matter scenario. Thus, below we will usually consider only fermion dark matter, but we emphasize that our results are approximately applicable to scalar dark matter as well. 

\begin{figure}
\center
	\includegraphics[width=0.495\textwidth]{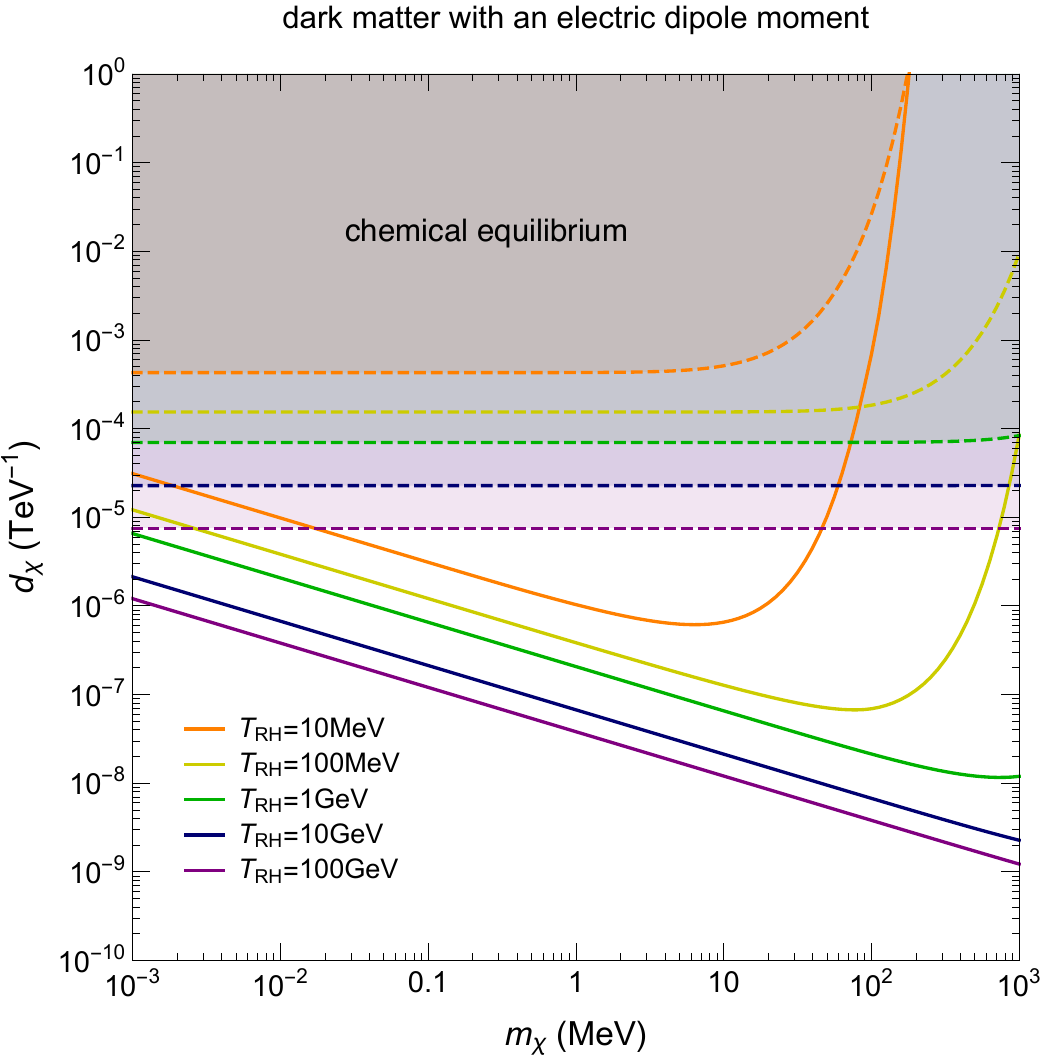}
	\includegraphics[width=0.495\textwidth]{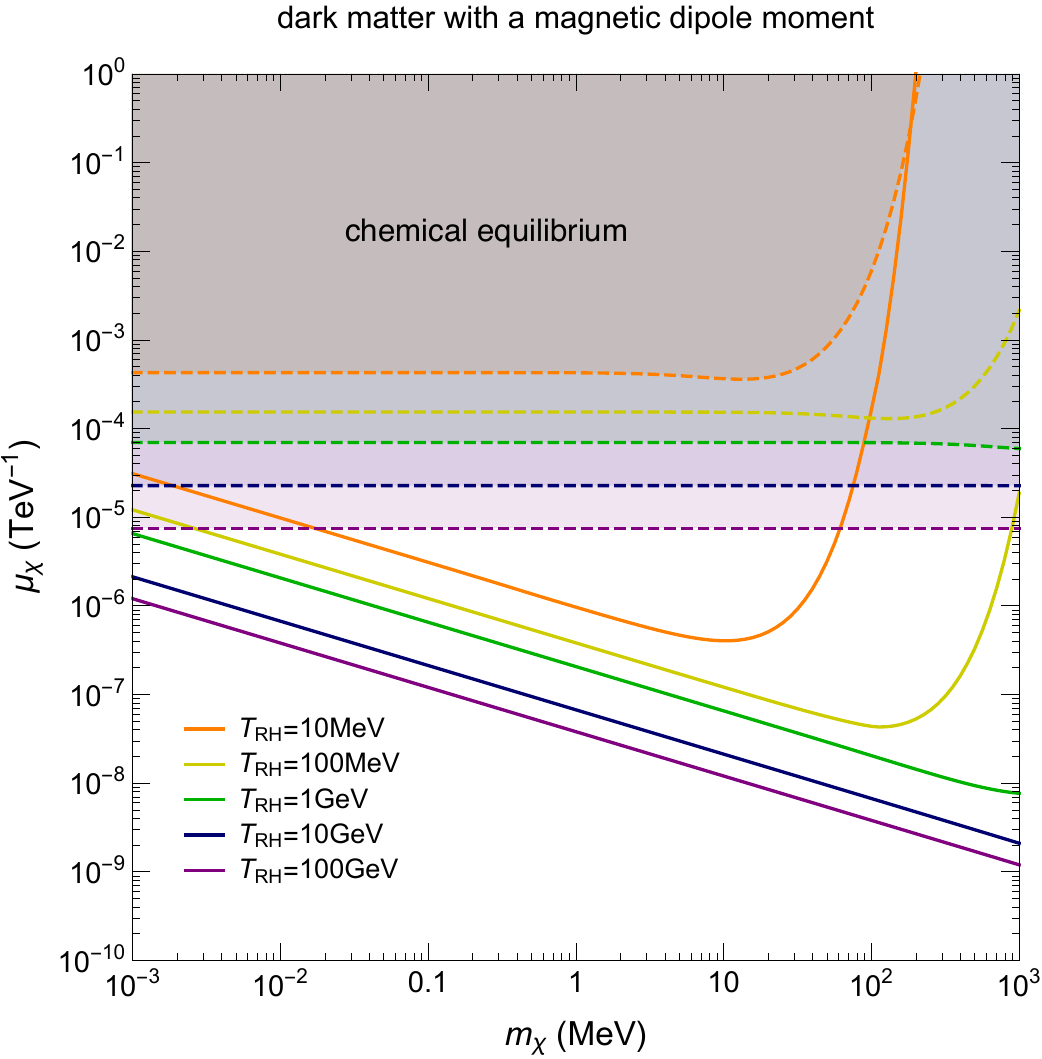}
	\caption{
	Solid lines in the left (right) plot show the values of the electric (magnetic) dipole moment needed to obtain the correct relic abundance from freeze-in for electric (magnetic) dipole dark matter for different reheating temperatures. 
For parameters above these ``freeze-in lines'', too much dark matter is produced in the early Universe. Above the dashed lines, the dark matter and SM sector are in chemical equilibrium, so that this parameter region is not compatible with any form of freeze-in production. 
Below the chemical equilibrium lines the model is safe from constraints on the number of relativistic degrees of freedom in the early Universe. 
}\label{fig:thermalDipoleM}
\end{figure}

\subsubsection{Dark matter with an Electric or Magnetic Dipole Moment}

The values of the electric (magnetic) dipole moment needed to obtain the correct relic abundance from freeze-in for electric (magnetic) dipole dark matter are shown for different reheating temperatures in the left (right) plot of Fig.~\ref{fig:thermalDipoleM}. Above the lines, too much dark matter is produced in the early Universe, whereas below the lines the relic abundance is lower than the observed amount of dark matter.
We find that the dark matter production via freeze-in is dominated for all $m_\chi$ by the pair-annihilation process rather than from plasmon decay.\footnote{We thank Shiuli Chatterjee and Ranjan Laha for helpful discussion.}  This is because production via plasmon decays is subdominant compared to production via pair annihilation at high temperatures, while  dark matter particles of electron and dipole moments are produced via UV freeze-in. 
%This is because the plasmon decay gives the largest contribution at $T \sim m_\chi$ compared to the pair annihilation \re{not clear?}. 
%Towards larger values of $m_\chi$, the production is dominated by the annihilation of SM particles. For reheat temperatures, $T_\text{RH}$, with $m_\chi\lesssim T_\text{RH}/20$, the production from plasmon decay dominates, giving rise to the kink in the freeze-in lines at this mass scale. The position of the kink is caused by the fact that the plasma frequency, which is a good estimate for the available phase space in the plasmon decay, is roughly $T_\text{RH}/10$ and the process starts at the reheating epoch (due to the high energy tail of the thermal distribution, the kink is near masses of $T_\text{RH}/10$ rather than $T_\text{RH}/20$). 
Dark matter with a mass above the reheating temperature cannot be produced efficiently in the early Universe and thus large values of the electric or magnetic dipole moments are needed.  However, these values become so large for increasing dark matter masses that the freeze-in line intersects the chemical equilibrium line, so that for even larger dark matter masses, the observed relic abundance cannot be obtained from freeze-in for any value of the dipole moment. 

\section{Stellar Constraints}\label{sec:stellarconstraints}
For dark matter masses below $\sim$100~keV, the models under consideration can be constrained from stellar cooling arguments~\cite{Raffelt:1996wa}. 
In some stages of stellar evolution the energy loss into a dark sector is severely constrained from astrophysical observations of globular clusters. We will briefly review these arguments and then discuss the resulting limits.

\subsection{Critical Stages of Stellar Evolution}\label{sec:StellarEvolution}
A globular cluster is a star cluster with a particularly high density of stars. It is tightly gravitationally bound and has a spherical shape. It is a satellite of a galaxy and most likely it was formed within the star formation process of the parent galaxy. 
Low mass stars in globular clusters can be used to constrain particle physics properties by the stars' characteristic properties of helium ignition and burning. 

To understand the origin of the constraints, it is useful to look at the Hertzsprung-Russel-diagram (HRD) of the globular cluster. Each single star is presented by a point in the plane spanned by the absolute brightness (in magnitudes) and the spectral classes. Note that the latter is correlated with the surface temperature, increasing from the right to the left.
Within its lifetime, a star moves through the diagram, starting on the so-called `main sequence', which is the diagonal from the lower right to the upper left corner. In this stage, it is burning hydrogen to helium in the core. Once the core is transformed into helium, the fusion process moves outwards, building a shell around the core. At this stage, the star becomes a red giant (RG), increasing its magnitude and moving to the red giant branch in the HRD. Depending on the mass of the star, it continues burning helium and eventually heavier elements in the horizontal branch or as a super giant. In its final stage, it becomes either a white dwarf, which is found in the lower left corner of the HRD (faint and hot), or a neutron star or black hole, which are not depicted, since they have no brightness.

\subsubsection{Helium Ignition in Red Giants}\label{subsec:RG}
For low mass stars ($0.5 \Ms \lesssim M \lesssim 2.3 \Ms$) helium ignition starts once the core has accumulated to roughly $0.5\Ms$.  At this stage of stellar evolution the star has reached the tip of the red giant branch. 
An extra source of cooling would delay helium ignition. The resulting heavier core would imply longer hydrogen burning in the shell and thus a brighter red giant. From the magnitude of the tip of the red giant branch one can thus constrain unknown elementary processes that would enhance the cooling of the star.

Simulations have shown that an extra energy loss of $\lesssim 10\ \text{erg g}^{-1}\text{s}^{-1}$~\cite{Raffelt:1996wa} is consistent with observations.
The core density of the red giant is on average $2\times 10^5\ \text{g/cm}^3$ and varies only within a factor of order one. The electrons are degenerate. The temperature is $10^8\ $K and the electron concentration is $Y_e=0.5$.

\subsubsection{Lifetime on the Horizontal Branch}
Once the star is burning helium it moves to the so-called `horizontal branch' (HB) in the HRD. The stars have a core mass of roughly $0.5\Ms$. The stars on the HB differ only in the mass of their hydrogen shell, and hence they have different surface temperatures (or spectral classes) but a similar magnitude; this is why in the HRD they lie on a horizontal line.  A globular cluster has hundreds of thousands of stars. This allows one to determine the lifetime of a star on the HB from the ratio of the number of stars on the HB to the number of stars on the red giant branch. It agrees with the prediction from the stellar standard models within $10\%$. However, an exotic contribution to cooling would result in a faster fuel consumption and has been constrained to be smaller than $10\ \text{erg g}^{-1}\text{s}^{-1}$~\cite{Raffelt:1996wa}. 
The temperature and electron concentration are the same as for the red giants discussed in Sec.~\ref{subsec:RG} but the density is slightly smaller, $0.6\times 10^4\ \text{g/cm}^3$. In this case, the electrons are not degenerate.

%\subsection{Solar Luminosity}\label{subsec:solar-lumi}
%Another constraint can be deduced directly from the minimal lifetime of the Sun. An extra efficient hydrogen burning mechanism would result in faster fuel consumption, reducing the Sun's overall lifetime.  Observations of the Sun's age today suggest that the energy emitted via the dark sector should not exceed the `normal' solar luminosity, $L_\odot=3.84\times 10^{33}$ erg/s. We take the temperature and density profile of the Sun from the model BS2005~\cite{Bahcall:2004pz}. The temperature in the core of the Sun reaches only $\sim 10^7$~K, which corresponds to roughly 1~keV. 

\subsection{Dark Matter Production Mechanisms In Stars}
The energy-loss rate or luminosity per unit volume in stellar objects can be written as
\begin{equation}
\label{eq:dLdV}
\frac{\dd L_\chi}{\dd V} = \int \dd \Pi_i \dd \Pi_f E_\textrm{out} |\mathcal{M}_{i \rightarrow f}|^2 (2\pi)^4 \delta^{(4)} \left( \Sigma P_i - \Sigma P_f \right).
\end{equation}
Here, we use the same notation as in Eq.~\eqref{eq:RT}, and $E_\textrm{out}$ is the energy sum of outgoing dark sector particles.

The electron-photon plasma inside stars gives rise to several production channels of light dark matter particles.
The core temperature of the stellar objects discussed above reaches up to $\sim$10~keV. In this energy regime, the dominant production channel for light dark matter particles is through plasmon decays. Additionally, for dark matter masses above the plasma frequency where the plasmon decay is kinematically suppressed, the production via bremsstrahlung processes is relevant. In the following, we will discuss the plasmon decay and the bremsstrahlung processes.  
The total luminosity will then be given by 
\begin{equation}
\frac{\dd L_\chi}{\dd V} = \frac{\dd L_\chi^\textrm{plasmon}}{\dd V}+\frac{\dd L_\chi^\textrm{brem}}{\dd V} \ .
\end{equation}
We note that these processes for the dark matter models with an electric and a magnetic dipole moment have been calculated in detail recently in~\cite{Chu:2019rok}, which appeared during the latter stages of completing our work.  
We have checked that the production via Compton scattering is subdominant to the production from bremsstrahlung and plasmon decays, so we do not consider it further.  We refer the reader to~\cite{Chu:2019rok} for the detailed calculations also for the production via Compton scattering.

\subsubsection{Plasmon Decays}
As discussed in Sec.~\ref{sec:plasmonMatrixEl}, the continuous interaction of photons with the free electrons in the plasma gives rise to quasi-massive longitudinal and transverse modes of the photon. The dispersion relations in the plasma (see Eqs.~\eqref{eq:dispersionT} and \eqref{eq:dispersionL} in Appendix~\ref{app:plasmaProp}), allow decays to massive particles like neutrinos or dark matter. 
An estimate of the maximum dark matter mass that can be produced from plasmon decays is given by the plasma frequency $\omega_p$ (see Eq.~\eqref{eq:omegaP}), which reaches roughly $1.6$~keV in stars on the horizontal branch and $8.6$~keV in red giants before helium ignition (for comparison, it is $0.3$~keV in the solar core). 

For plasmon decays, Eq.~\eqref{eq:dLdV} can be written as~\citep{Chang:2018rso} 
\begin{align}\label{eq:dLdVdecay}
\frac{\dd L_\chi^\textrm{plasmon}}{\dd V}=\int \frac{\dd^3k}{(2\pi)^3}\left(\frac{2\omega_T\Gamma_T}{e^{\omega_T/T}-1}+\frac{\omega_L\Gamma_L}{e^{\omega_L/T}-1}\right)\ .
\end{align}
The form of this formula is simple to understand: the total energy carried away in the dark sector is given by the energy of the decaying plasmon $\omega_{T,L}$ and the rate with which it decays, $\Gamma_{T,L}$. The factor of two in the numerator of the first term accounts for the two transverse modes, and the denominator comes from the Bose-Einstein-distribution that is obeyed by the photons in the star of temperature $T$.
The plasmon decay rate to the dark sector is given by Eq.~\eqref{eq:plasmondecayrate}
%\begin{align}\label{eq:starGammaTL}
%\Gamma_{T,L}&=\frac{1}{2\omega_{T,L}}\int\frac{\dd^3p_\text{DM}}{(2\pi)^3 2E_\text{DM}}\frac{\dd^3p_{\overline{\text{DM}}}}{(2\pi)^3 2E_{\overline{\text{DM}}}}(2\pi)^4\delta^4\left(K-P_\text{DM}-P_{\overline{\text{DM}}}\right)\left|\mathcal{M}_{T,L}\right|^2\\
%&=\frac{1}{16\pi\omega_{T,L}}\int\dd\cos\theta\frac{p_\text{DM}^2}{E_\text{DM} E_{\overline{\text{DM}}}}\left(\frac{\dd g(k,\theta)}{\dd p_\text{DM}} \right) ^{-1}\left|\mathcal{M}_{T,L}\right|^2\ ,
%\end{align}
%where again $\theta$ is the angle between the incoming photon and outgoing DM, and $K_{T,L}=(\omega_{T,L},\vec{k})$ is the four-vector of the transverse or longitudinal plasmon, respectively.  Also, $E_{\overline{\text{DM}}}=\sqrt{k^2+m_\text{DM}^2+p_\text{DM}^2-2kp_\text{DM}\cos\theta}$, and $p_\text{DM}$ is given by the solution of $g(k,\theta)=0$ with
%\begin{align}
%g(k,\theta)=\omega(k)-\sqrt{p_\text{DM}^2+m_\chi^2}-\sqrt{k^2+p_\text{DM}^2-2kp_\text{DM}\sin\theta+m_\text{DM}^2}\ .
%\end{align}

The difference between the rate derived here compared to the freeze-in rate Eq.~\eqref{eq:ratePlasmon} is only the factor of the energy $\omega_{T/L}$ in Eq.~\eqref{eq:dLdVdecay}. To derive the stellar constraints, we are interested in the energy that is taken away from the star, i.e., the energy of the decaying plasmon.  For the freeze-in production in the early Universe, the relevant quantity is the number of particles that go into the dark sector. 

%The only model-dependent quantity that enters into the computation is the matrix element in Eq.~\eqref{eq:starGammaTL}.  These are given in Sec.~\ref{sec:plasmonMatrixEl} for the models discussed in this paper.

\begin{figure}[t]
\center
\begin{tikzpicture}[line width=1.3 pt, scale=2]
	\draw[fermion] (0,0.4)--(0.4,0.4);
	\draw[fermion] (0.4,0.4)--(0.8,0.4);
	\draw[fermion] (0.8,0.4)--(1.6,0.4);
	\draw[vector] (0.8,0.4)--(0.8,-0.43);
	\draw[fermion] (0,-0.43)--(0.8,-0.43);
	\draw[fermion] (0.8,-0.43)--(1.6,-0.43);
	\draw[vector] (0.4,0.4)--(0.58,0.544);
	\draw[vector] (0.72,0.656)--(0.9,0.8);
    	\draw[fermion] (0.9,0.8)--(1.4,1.);
    	\draw[fermionbar] (0.9,0.8)--(1.4,0.6);

	\node at (0.65, .6) {$\bigotimes$};
	\node at (-.15,.4) {$e^{-}$};
	\node at (-.15, -.43) {$p$};
	\node at (1.75,.4) {$e^{-}$};
	\node at (1.75, -.43) {$p$};
	\node at (0.4,0.6) {$\gamma$};
	\node at (0.75,0.85) {$A'$};
	\node at (1.55,1) {$\chi$};
	\node at (1.55,.6) {$\bar\chi$};

	\draw[fermion] (4,0.4)--(4.8,0.4);
	\draw[fermion] (4.8,0.4)--(5.2,0.4);
	\draw[fermion] (5.2,0.4)--(5.6,0.4);
	\draw[vector] (4.8,0.4)--(4.8,-0.43);
	\draw[fermion] (4,-0.43)--(4.8,-0.43);
	\draw[fermion] (4.8,-0.43)--(5.6,-0.43);
	\draw[vector] (5.2,0.4)--(5.38,0.544);
	\draw[vector] (5.52,0.656)--(5.7,0.8);
    	\draw[fermion] (5.7,0.8)--(6.2,1.);
    	\draw[fermionbar] (5.7,0.8)--(6.2,0.6);

	\node at (5.45, .6) {$\bigotimes$};
	\node at (3.75,.4) {$e^{-}$};
	\node at (3.75, -.43) {$p$};
	\node at (5.75,.4) {$e^{-}$};
	\node at (5.75, -.43) {$p$};
	\node at (5.2,0.6) {$\gamma$};
	\node at (5.55,0.85) {$A'$};
	\node at (6.35,1) {$\chi$};
	\node at (6.35,.6) {$\bar\chi$};
\end{tikzpicture}

\vspace{1 cm}

\begin{tikzpicture}[line width=1.3 pt, scale=2]
	\draw[fermion] (0,0.4)--(0.4,0.4);
	\draw[fermion] (0.4,0.4)--(0.8,0.4);
	\draw[fermion] (0.8,0.4)--(1.6,0.4);
	\draw[vector] (0.8,0.4)--(0.8,-0.43);
	\draw[fermion] (0,-0.43)--(0.8,-0.43);
	\draw[fermion] (0.8,-0.43)--(1.6,-0.43);
	\draw[vector] (0.4,0.4)--(0.9,0.8);
    	\draw[fermion] (0.9,0.8)--(1.4,1.);
    	\draw[fermionbar] (0.9,0.8)--(1.4,0.6);
	
	\draw[fill=black] (0.9,0.8) circle (.1cm);
	\draw[fill=white] (0.9,0.8) circle (.098cm);
	\begin{scope}
    	\clip (0.9,0.8) circle (.1cm);
    	\foreach \x in {.65,.75,...,4.05}
			\draw[line width=1 pt] (\x,.7) -- (\x+.3,.9);
  	\end{scope}	

	\node at (-.15,.4) {$e^{-}$};
	\node at (-.15, -.43) {$p$};
	\node at (1.75,.4) {$e^{-}$};
	\node at (1.75, -.43) {$p$};
	\node at (0.55,0.7) {$\gamma$};
	\node at (1.55,1) {$\chi$};
	\node at (1.55,.6) {$\bar\chi$};

	\draw[fermion] (4,0.4)--(4.8,0.4);
	\draw[fermion] (4.8,0.4)--(5.2,0.4);
	\draw[fermion] (5.2,0.4)--(5.6,0.4);
	\draw[vector] (4.8,0.4)--(4.8,-0.43);
	\draw[fermion] (4,-0.43)--(4.8,-0.43);
	\draw[fermion] (4.8,-0.43)--(5.6,-0.43);
	\draw[vector] (5.2,0.4)--(5.7,0.8);
    	\draw[fermion] (5.7,0.8)--(6.2,1.);
    	\draw[fermionbar] (5.7,0.8)--(6.2,0.6);
	
	\draw[fill=black] (5.7,0.8) circle (.1cm);
	\draw[fill=white] (5.7,0.8) circle (.098cm);
	\begin{scope}
    	\clip (5.7,0.8) circle (.1cm);
    	\foreach \x in {5.05,5.15,...,7.05}
			\draw[line width=1 pt] (\x,.7) -- (\x+.3,.9);
  	\end{scope}		

	\node at (3.75,.4) {$e^{-}$};
	\node at (3.75, -.43) {$p$};
	\node at (5.75,.4) {$e^{-}$};
	\node at (5.75, -.43) {$p$};
	\node at (5.35,0.7) {$\gamma$};
	\node at (6.35,1) {$\chi$};
	\node at (6.35,.6) {$\bar\chi$};
\end{tikzpicture}

\caption{The diagrams contributing to the bremsstrahlung process of producing dark matter in stars. A radiated photon during the electron-proton collision can produce dark matter either through mixing with the dark photon (top two diagrams) or through the electric or magnetic dipole moment (bottom two diagrams).}\label{fig:FeynBrem} 
\end{figure}
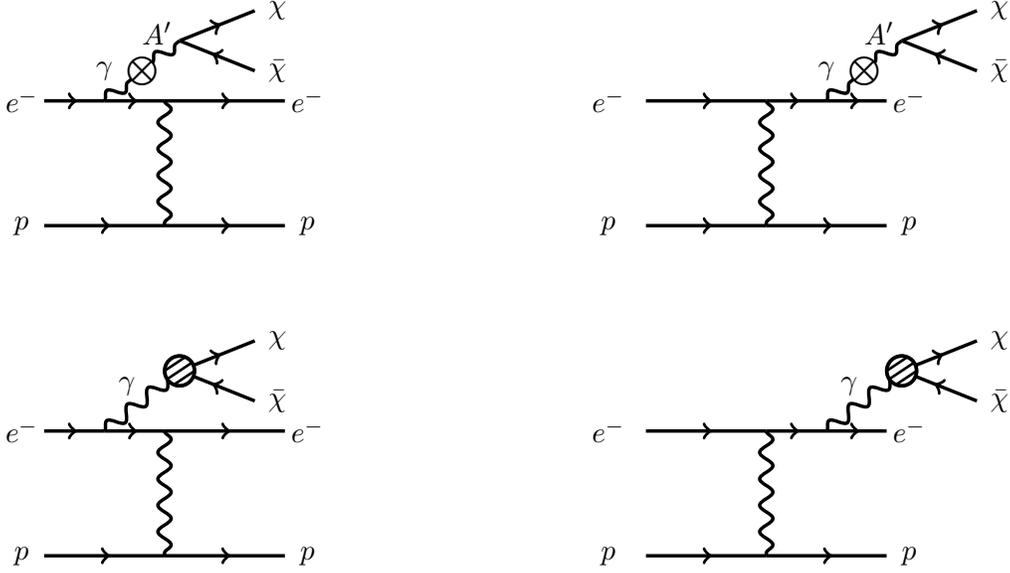

\subsubsection{Bremsstrahlung processes}\label{subsubsec:brem}
Dark matter particles can be produced by bremsstrahlung processes during the electron-proton collisions, see Fig.~\ref{fig:FeynBrem}.  For dark matter masses above the plasma frequency where the plasmon decay is forbidden, this process dominates. For the calculation of the bremsstrahlung processe, we use the soft radiation approximation (SRA).  Then, the energy loss rate Eq.~\eqref{eq:dLdV} for the bremsstrahlung can be separated into the collision part and the radiation part \cite{Chang:2018rso}.  Although the electron-proton collision is through a $t$-channel photon, which can have arbitrarily small momentum, the SRA is also valid because of the effective photon mass. Compared to the exact calculations in~\cite{Chu:2019rok}, the SRA gives the equivalent results up to an $O(1)$ factor.  For the process $e(P_1) + p (P_2) \rightarrow e(P_3) + e(P_4) + \chi (P_\chi) + \bar\chi (P_{\bar\chi}$), the rate is found to be
\begin{eqnarray}\label{eq:bremrate}
\frac{\dd L_\chi^\textrm{Brems}}{\dd V} &=& \int \dd \Pi_1 \dd \Pi_2 \dd \Pi_3 \dd \Pi_4 (2 \pi)^4 \delta^4(P_1+P_2-P_3-P_4) |\mathcal{M}|_{ep}^2  \frac{|\vec{p}_1-\vec{p}_3|^2}{m_e^2} \nonumber\\
&& \times \int \dd \Pi_k \dd s_\chi e^{-\omega/T} \frac{\alpha}{4 \pi (s_\chi+\omega_p^2)^2} \frac{3 \omega^2-k^2}{3 \omega^3} \sqrt{1-\frac{4 m_\chi^2}{s_\chi}} f_\text{DM}(s_\chi),
\end{eqnarray}
where the index $k$ corresponds to the radiated off-shell photon with 4-momentum $K=(\omega,\vec{k})$, $s_\chi = \omega^2 -k^2$, and
\begin{equation}
|\mathcal{M}|_{ep}^2 = 32\pi^2 \alpha^2 \frac{2(s-m_e^2-m_p^2)-2 s t -t^2}{t^2 - \omega_p^2} 
\end{equation}
is the electron-proton elastic scattering amplitude. See Appendix.~\ref{app:Brems} for detailed calculations. Here, we use the approximation that the self energy of the photon is $\omega_p^2$. Note we have used a plus sign in the s-channel propagator to avoid the double counting from the resonance. To get the second line in Eq.~\eqref{eq:bremrate}, we again follow the calculations in~\cite{Chu:2019rok} for integrating over the outgoing dark matter phase space.  The factor $f_\text{DM}(s)$ for each case is shown in Eqs.~\eqref{eq:fDMDP}-\eqref{eq:fDMMDM}.

\subsection{Results}\label{subsec:results}
\begin{figure}
\center
	\includegraphics[width=0.8\textwidth]{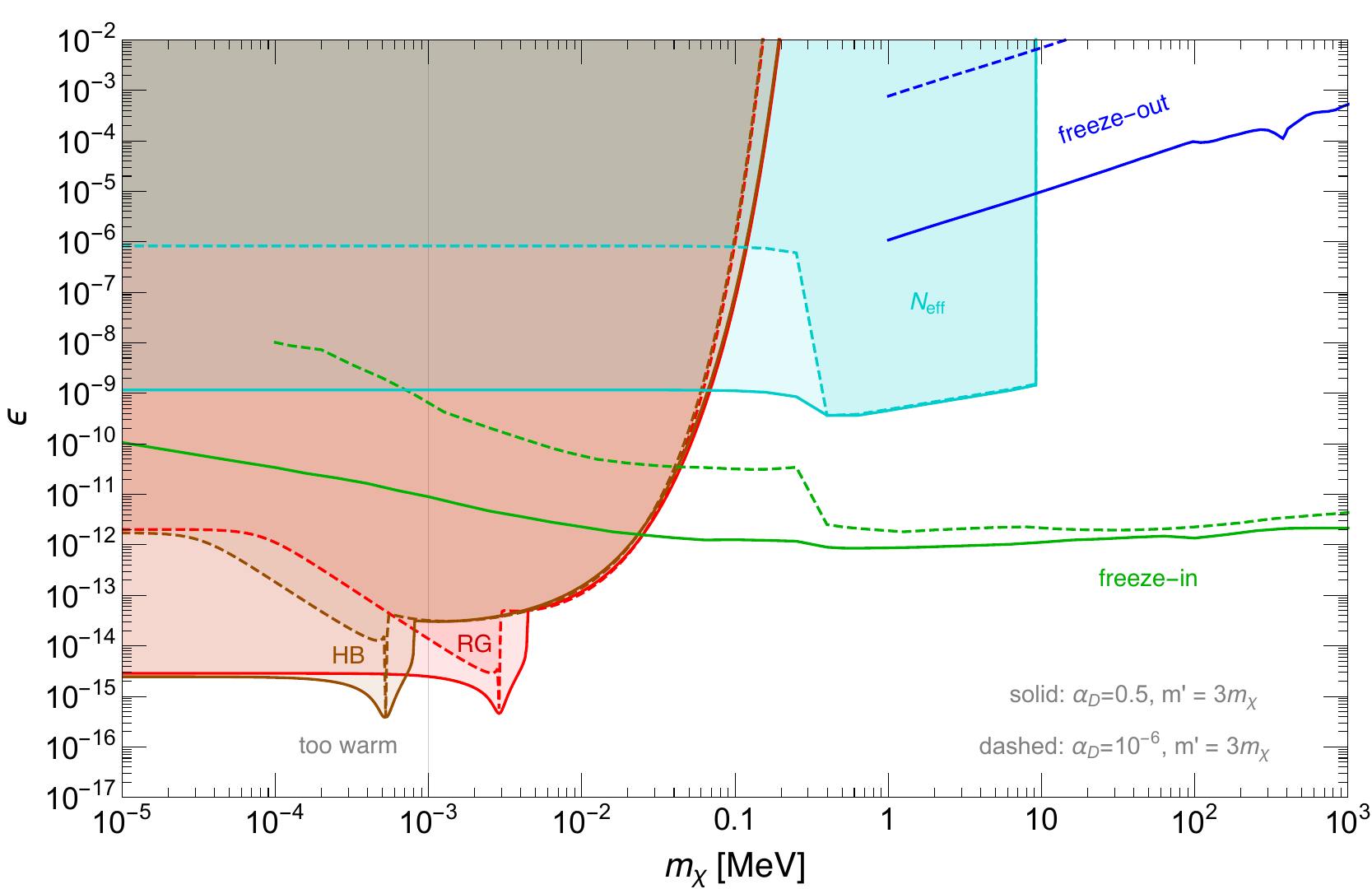}
	\caption{Stellar cooling constraints derived in this work on Dirac fermion dark matter interacting with a dark photon with dark photon masses $m^\prime=3m_\chi$ and $\alpha_D=0.5$ ($\alpha_D=10^{-6}$) for the solid (dashed) lines.  The cooling constraints are derived for stars on the horizontal branch (brown) and red giants (red).  In green, we show the parameters for which freeze-in production provides the entire dark matter relic abundance (see also Fig.~\ref{fig:FIfermion}); above the line too much dark matter would have been produced.  In blue, we show the parameters for which thermal freeze-out production provides the entire dark matter relic abundance.  
Above the cyan lines, the dark sector was in chemical equilibrium with the SM bath and is constrained below $m_\chi=9.4$~MeV by $N_\text{eff}$.  Below $\sim$1~keV dark matter is constrained from structure formation.  Other relevant constraints and some projections from terrestrial searches are shown in Fig.~\ref{fig:summaryHP}.  
The bounds on scalar dark matter coupling to a dark photon (not shown) are similar. 
}
\label{fig:stellarConsHP}

\end{figure}
The total dark luminosity is found by integrating Eq.~\eqref{eq:dLdVdecay} and Eq.~\eqref{eq:bremrate} over the volume of the star and summing the two contributions. If the density and temperature profile is non-trivial, like in the solar case, all quantities depend on the radius, which has to be taken into account for the spatial integration. The constraint on $\epsilon$ is found by requiring that the dark luminosity does not exceed the limits discussed in Secs.~\ref{sec:StellarEvolution}.

In Fig.~\ref{fig:stellarConsHP}, we show the stellar constraints for the dark photon portal dark matter for $m^\prime=3m_\chi$ and for $\alpha_D=0.5$ (solid lines) and $\alpha_D=10^{-6}$ (dashed lines).  The brown and red contours show the constraints from the lifetime on the horizontal branch (`HB') and the non-delay of helium ignition in red giants (`RG'), respectively. The plasma frequency in red giant stars is the highest, hence it can probe the largest dark matter masses.
When the plasma frequency equals the dark photon mass, the propagator in the cross section is on resonance. 
The production is enhanced at this parameter point, and the constraint is thus particularly strong.  This is seen in the spike-like features in the stellar constraints.
While we show the results for a dark matter fermion only, we again note that the bounds on scalar dark matter coupled to a dark photon are very similar. 

For the dark-photon-mediated dark matter, we compare in Fig.~\ref{fig:stellarConsHP} the stellar constraints to the freeze-in lines (in green), which are also shown in Fig.~\ref{fig:FIfermion}.  We find that dark matter that is entirely produced from this mechanism is ruled out below $\sim$20~keV for $\alpha_D=0.5$, and below $\sim$40~keV for $\alpha_D=10^{-6}$. 
Note that the areas between the respective freeze-in and freeze-out lines (blue) are forbidden in this model, as an overabundance of dark matter would have been produced, overclosing the Universe.  Additional decay modes (of the dark photon) beyond the ones assumed in the minimal model setup discussed here, or slight model variations, could open up some of this parameter region (see, e.g.,~\cite{Hochberg:2014dra,Kuflik:2015isi,Izaguirre:2015yja,DAgnolo:2015ujb,Pappadopulo:2016pkp,Farina:2016llk,DAgnolo:2017dbv,DAgnolo:2018wcn,DAgnolo:2019zkf,Hambye:2019dwd,Heeba:2019jho,Battaglieri:2017aum,Evans:2019vxr}). 

Note that we have not derived the constraints from the cooling of white dwarfs and the sun. In the high-mass regime where the bremsstrahlung processes dominate, it is not competitive with the other stellar cooling constraints as the white dwarfs and the sun have a much lower temperature than red giant stars. However, due to the high density the white dwarfs have a high plasma frequency, of $\sim$23~keV. Thus, a small fraction of the parameter space on the right-hand-side of the red giant tip can in principle be excluded additionally (see e.g.~\cite{Vogel:2013raa}, where this was shown for dark photon dark matter). 

\begin{figure}
\center
	\includegraphics[width=0.495\textwidth]{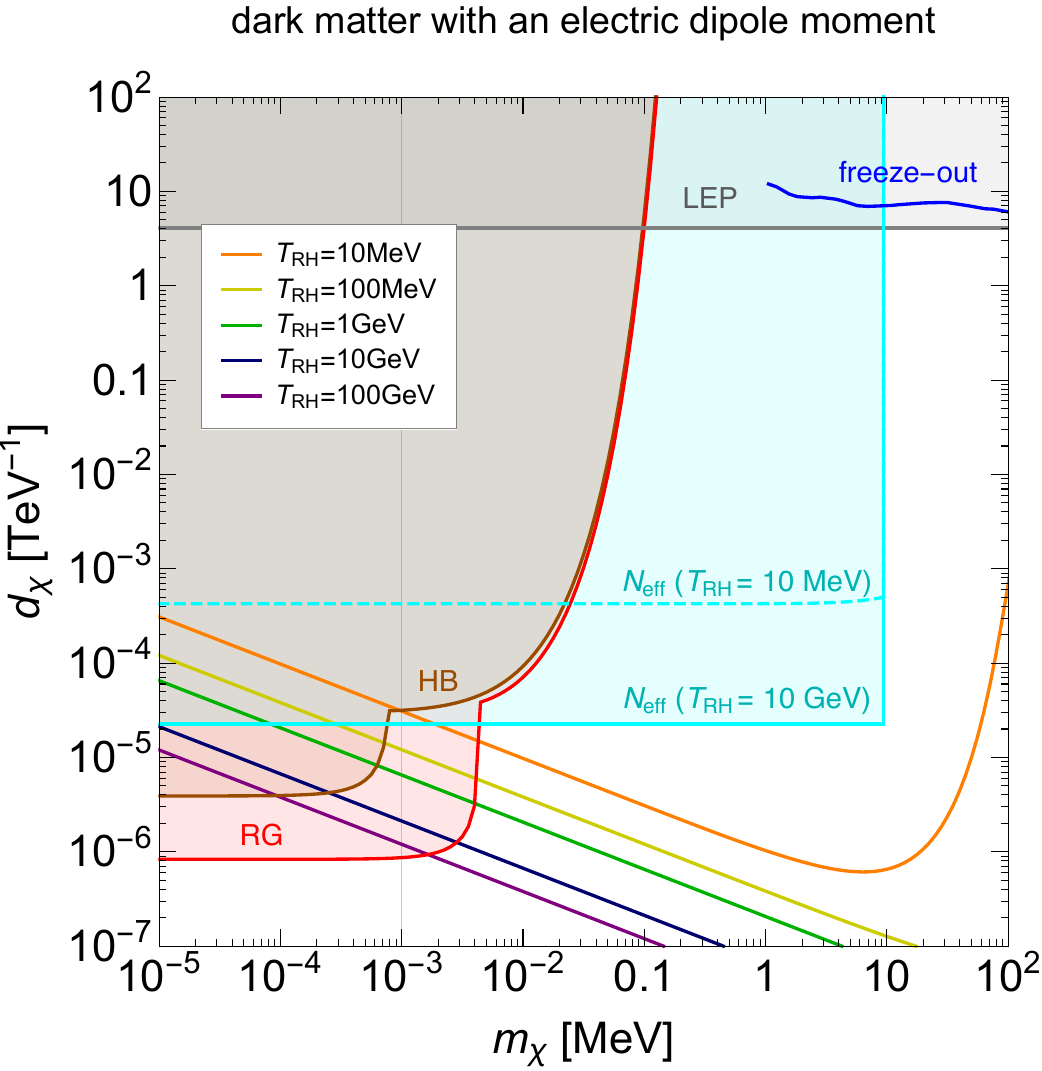}
	\includegraphics[width=0.495\textwidth]{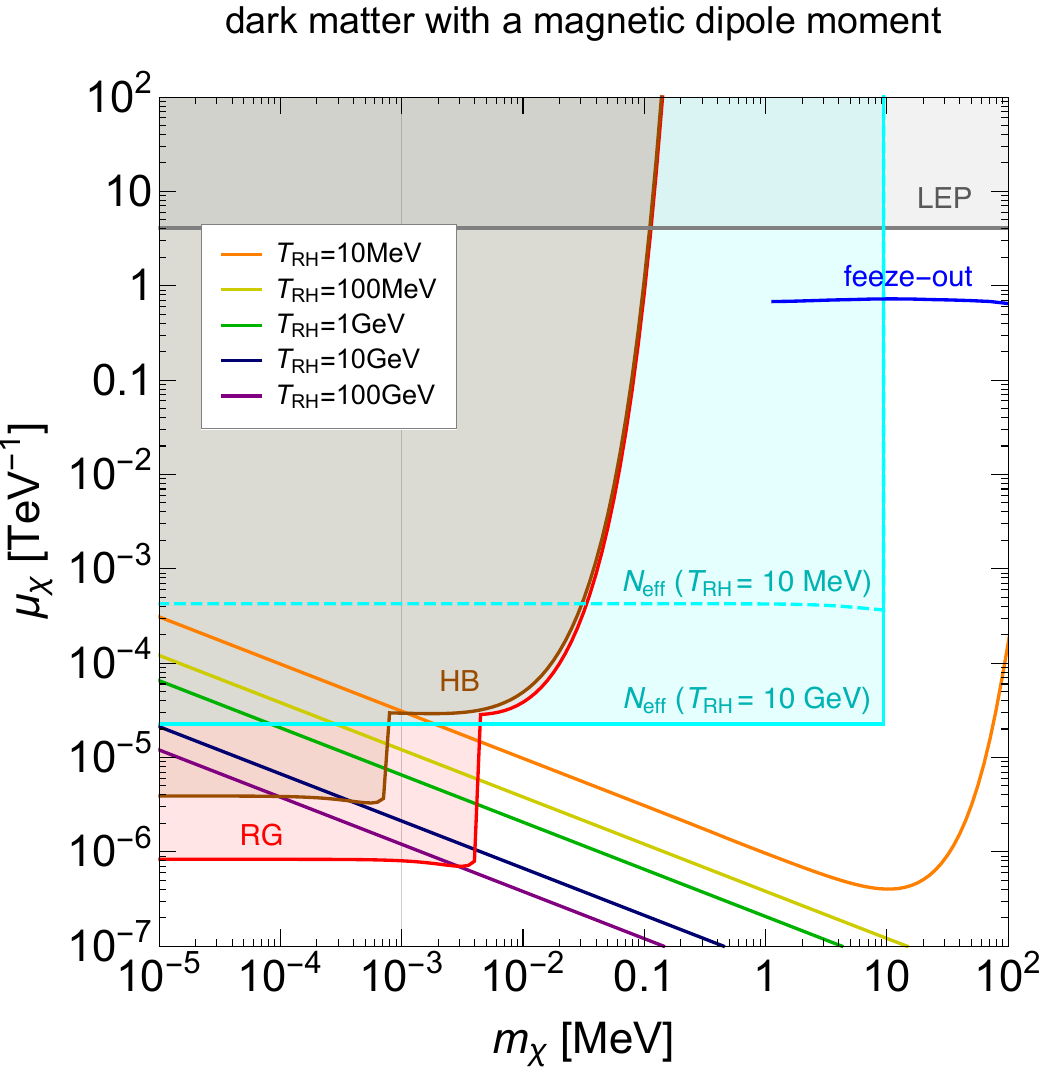}
	\caption{
Stellar cooling constraints derived in this work on dark matter with an electric dipole moment (left) or a magnetic dipole moment (right), from stars on the horizontal branch (brown) and red giants (red). 
	We also show lines for different reheating temperatures along which freeze-in production provides the entire dark matter relic abundance (see also Fig.~\ref{fig:thermalDipoleM}); above the line too much dark matter would have been produced.  Above the cyan lines, the dark sector was in chemical equilibrium with the SM bath and is constrained below $m_\chi=9.4$~MeV by $N_\text{eff}$. 
Below $\sim 1$ keV dark matter is constrained from structure formation. Above the gray line the models are constrained from LEP data.  The blue curve shows the parameters needed to obtain the correct relic abundance from thermal freeze-out. 
}\label{fig:stellarConsEDM}
\end{figure}

The stellar constraints for dark matter with a dipole moment are shown in Fig.~\ref{fig:stellarConsEDM}, together with the freeze-in lines. The left (right) plot shows the limits for dark matter with an electric (magnetic) dipole moment, respectively. The limits from the red giants are always stronger than the ones from the horizontal branch stars. 
No resonant production occurs due to the absence of a mediator in that mass range. We find that for dark matter with a dipole moment, the freeze-in mechanism is constrained for $m_\chi \lesssim 2 \-- 5$ keV depending on the reheating temperature. 
We show also the freeze-out parameters from~\cite{Sigurdson:2004zp}, as well as the LEP limit from~\cite{Fortin:2011hv}. 

The stellar constraints do not have upper boundaries, unlike the supernova 1987A constraints~\cite{Chang:2018rso,Raffelt:1996wa}. 
Consider first the supernova 1987A constraint.  In this case, the lower boundary of the constrained region is set by the requirement of producing a sufficient number of dark sector particles to carry away more energy than that carried away by neutrinos, which are believed to dominate the energy loss.  This would drastically change the cooling of the proto-neutron star, in conflict with observations.  The upper boundary of the constrained region arises from a sufficient number of dark sector particles becoming trapped, thermalizing with the matter inside the proto-neutron star, and failing to carry away sufficient energy.  Roughly speaking, if the dark sector particles couple more strongly than neutrinos to the matter inside the proto-neutron star (mostly protons, neutrons, and electrons), the dark sector particles are unable to carry away enough energy, and there is no constraint.  This is why there is no supernova 1987A constraint up to arbitrarily high couplings. However, in the case of stellar cooling, the photon dominates the energy loss of the stars.  
Since it is impossible for the dark sector particles considered in this paper to have stronger couplings to photons than SM particles, 
the dark sector particles will always carry away more energy than the photon.  Moreover, the criteria used for the stellar cooling bounds is that the dark sector particles must carry away \textit{less} than a fraction of the energy carried away by photons.  Thus, only if the dark sector particles interact more strongly than photons, would they fail to carry away sufficient energy.  Therefore, there is no upper boundary for the stellar cooling constraints. 

\section{Potential Reach of Terrestrial Searches}\label{sec:lab}

\begin{figure}[t]
\center
	\includegraphics[width=0.8\textwidth]{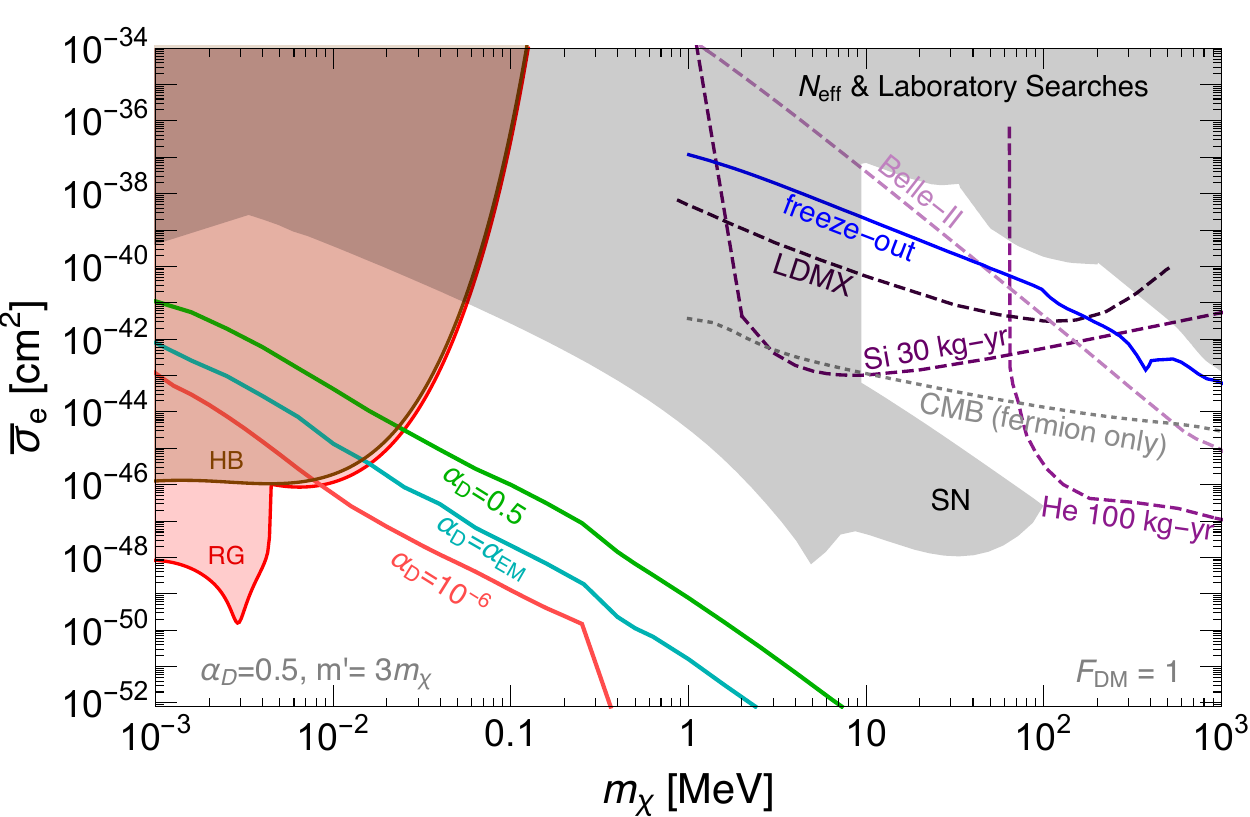}
	\caption{
Solid lines in green, cyan, and red show the values of the dark-matter-electron-scattering cross section for which the correct dark matter relic abundance is obtained from freeze-in for Dirac fermion dark matter coupled to a dark photon, for various choices of the 
dark-photon couplings $\alpha_D$ (see Fig.~\ref{fig:FIfermion}). Red and brown-shaded regions show the stellar constraints from red giant and horizontal branch stars (see Fig.~\ref{fig:stellarConsHP}). 
Dashed lines show the potential reach of laboratory experiments. The gray shaded regions are excluded from the number of effective relativistic degrees of freedom (see Fig.~\ref{fig:stellarConsHP}), supernova 1987A, and existing laboratory constraints. The dotted line shows the CMB constraint, which excludes the freeze-out line (blue) when the dark matter particle is a Dirac fermion.  Projections and constraints for dark matter that is a scalar particle are similar, except with a much weaker CMB bound.  
	If not stated otherwise, the model parameters are $m^\prime=3m_\chi$ and $\alpha_D=0.5$.  See text for details. 
	}\label{fig:summaryHP}
\end{figure}

\begin{figure}[t]
\center
	\includegraphics[width=0.495\textwidth]{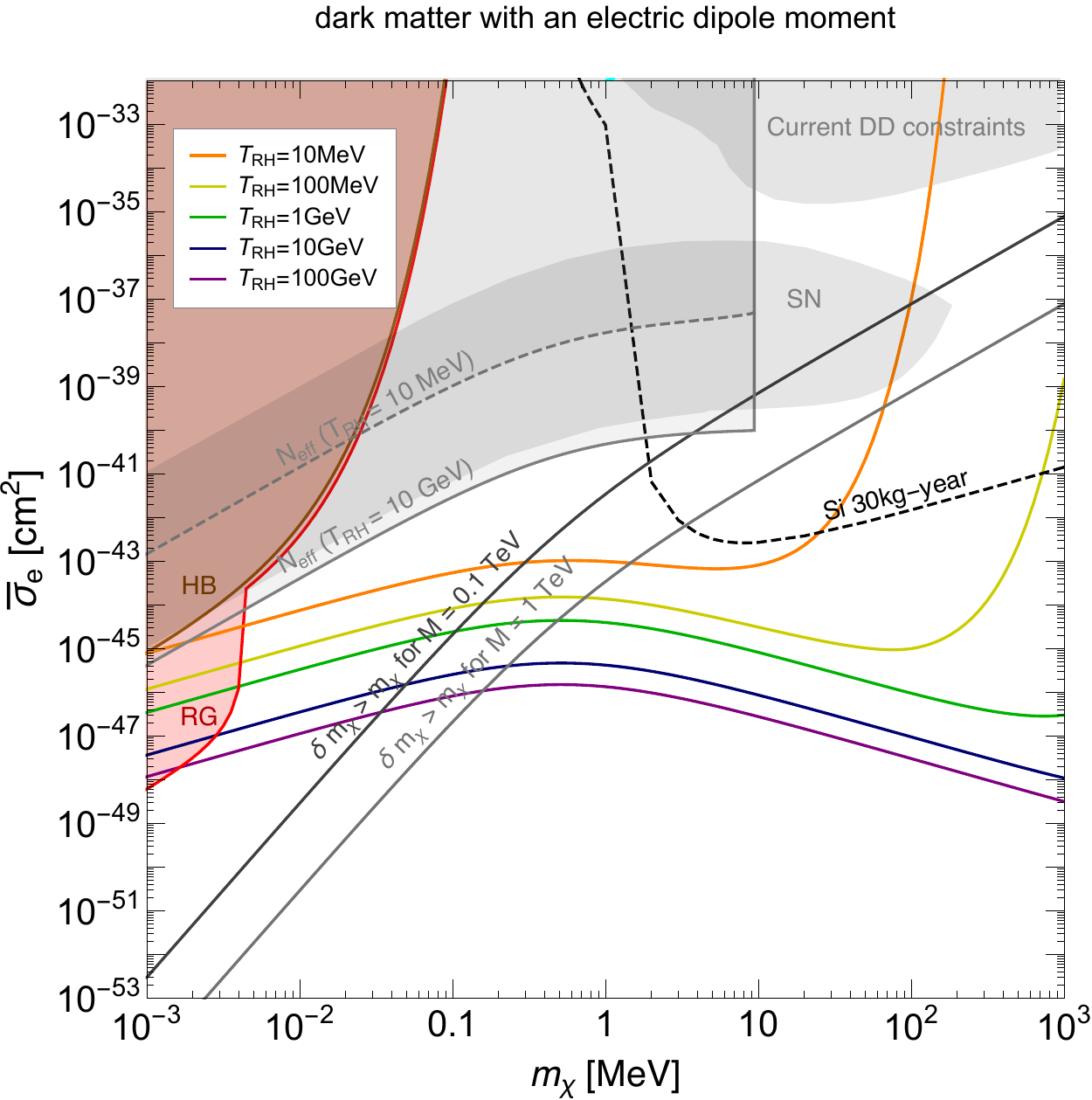}
	\includegraphics[width=0.495\textwidth]{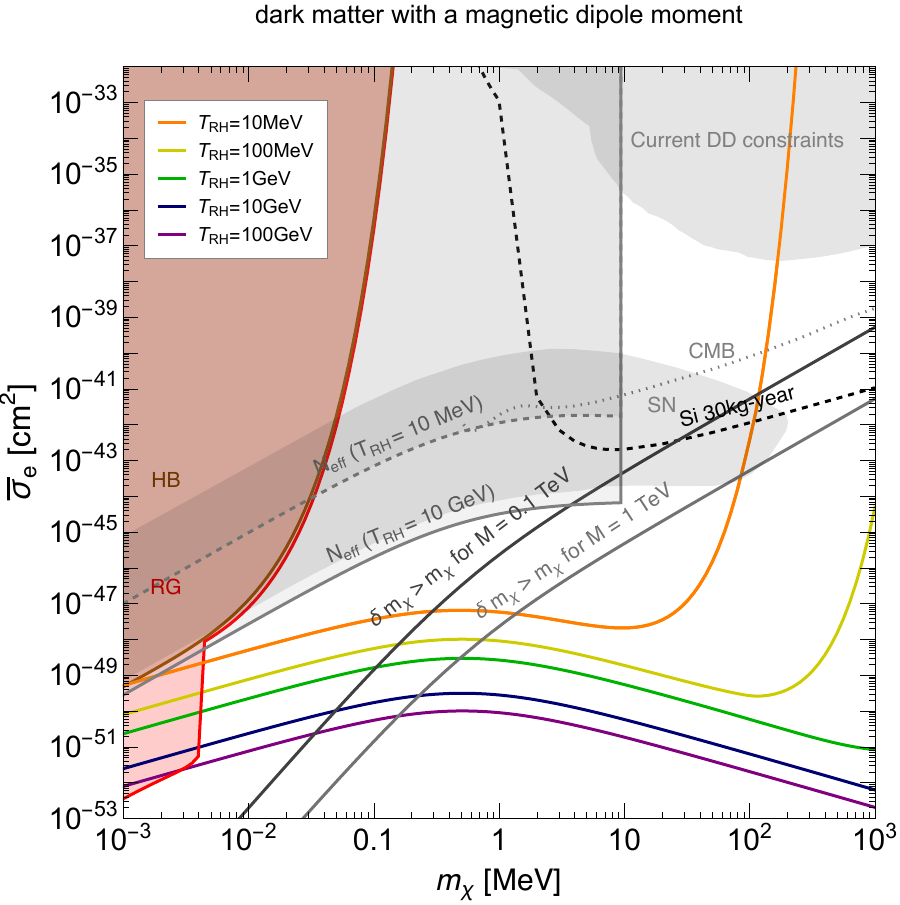}
	\caption{
	Solid colored lines show the values of the dark-matter-electron-scattering cross section for which the correct dark matter relic abundance is obtained from freeze-in for dark matter interacting with an electric (left) or magnetic (right) dipole moment, for various reheating temperatures (see Fig.~\ref{fig:thermalDipoleM}). Red and brown-shaded regions show the stellar constraints from red giant and horizontal branch stars (see Fig.~\ref{fig:stellarConsEDM}). 
The dashed line shows the potential reach of a direct-detection experiment using Skipper-CCDs for a 30~kg-year exposure. 
The gray shaded areas are excluded from the number of effective relativistic degrees of freedom (see Fig.~\ref{fig:stellarConsEDM}), direct detection searches, and supernova 1987A.  For the right plot, the region above the dotted line is excluded from the CMB.  The freeze-in line for $T_\text{RH}=10$~MeV (orange) stops at the coupling where dark matter would thermalize with the SM sector. 
See text for details. 
}
\label{fig:summaryEDM}
\end{figure}

The freeze-in dark matter models discussed in this paper are challenging to detect in the laboratory with direct-detection and accelerator-based experiments.  This is not surprising, given the small required couplings.  Nevertheless, we illustrate this challenge in Fig.~\ref{fig:summaryHP} for a ``heavy'' dark photon mediator with $m'=3m_\chi$ and $\alpha_D=0.5$ (unless otherwise indicated), and in Fig.~\ref{fig:summaryEDM} for dark matter interacting with an electric or magnetic dipole moment.  We parameterize as usual the reference dark-matter-electron scattering cross section, $\overline{\sigma}_e$, and form factor for the dark matter, $ |F_{\rm{DM}}(q)|^2$, as~\cite{Essig:2011nj,Essig:2015cda} 
\bea
\overline{|{\cal{M}}_{\rm{free}}(\vec q)|^2}&\equiv&\overline{|{\cal{M}}_{\rm{free}}(\alpha m_e)|^2}\times  |F_{\rm{DM}}(q)|^2 
\label{eq:sigmaebar} \\
\overline\sigma_e&\equiv&\frac{\mu_{\chi e}^2\overline{|{\cal{M}}_{\rm{free}}(\alpha m_e)|^2}}{16\pi m_\chi^2 m_e^2}, \label{eq:FDM} 
\eea
where $\overline{|{\cal{M}}_{\rm{free}}|^2}$ is the absolute value squared of the elastic dark-matter-(free)-electron matrix element and $q$ is the magnitude of the three-momentum lost by the dark matter when it scatters off the electron. 
For each of these models, we can derive $\bar\sigma_e$ to be 
\begin{eqnarray}
\overline\sigma_e^{A^\prime} &= &  \frac{16 \pi \alpha \alpha_D \epsilon^2 \mu_{\chi e}^2}{(\alpha^2 m_e ^2+m^{\prime 2})^2} \\
\overline\sigma_e^{\text{EDM}} &= &  \frac{4 d_\chi^2 \mu_{\chi e}^2}{\alpha m_e^2}\\
\overline\sigma_e^{\text{MDM}} & = & \frac{\alpha \mu_\chi^2 \mu_{\chi e}^2}{m_e^2 m_\chi} \left(m_\chi-2m_e +\frac{4 m_\chi^2 v_\text{rel}^2}{\alpha^2} \right) \simeq \frac{5 \alpha \mu_\chi^2 \mu_{\chi e}^2}{m_e^2} ~~(m_\chi\gg m_e,~v_\text{rel}\simeq \alpha)\ ,\label{eq:sigmaemdm}
%& \simeq & \frac{5 \alpha \mu_\chi^2 \mu_{\chi e}^2}{m_e^2}\ ,  \qquad m_\chi\gg m_e,~v_\text{rel}\simeq \alpha\ ,
\end{eqnarray}
where $\mu_{\chi e}$ is the reduced mass between the electron and $\chi$, and $v_\text{rel}$ is the relative velocity between the incoming dark matter and the incoming electron. For simplicity, we use the approximation in Eq.~\eqref{eq:sigmaemdm} even for $m_\chi < m_e$.

A dark photon mediator can be classified as ``heavy'' and give $F_{\rm DM}=1$ once its mass is above the typical momentum transfer, $q_\text{typ}$, which  varies for different targets.  For example, for direct-detection experiments with semiconductor or noble liquid targets, $q_\text{typ} \equiv \mu_{\chi,e} v_\text{rel} \simeq \alpha m_e$~\cite{Essig:2015cda}.  So for dark photon masses above a few keV (which is enforced by the stellar constraints), we have $F_{\rm DM}=1$.    
Dark matter interacting with an electric dipole moment has the form factor 
%$F_{\rm DM}=\alpha m_e/q$, where $q$ is the momentum transfer. The form factor for dark matter interacting with a magnetic dipole moment is 
\begin{equation}
F_{\rm DM}=\alpha m_e/q \qquad\qquad \text{(EDM)}\, .
\end{equation}  
The form factor for dark matter interacting with a magnetic dipole moment is more complicated, 
\begin{eqnarray}
F^2_{\text{DM}}(q)& \simeq & \frac{1}{(5m_\chi-2m_e)}\left((m_\chi-2m_e) + \frac{4 m_e^2 m_\chi v_\text{rel}^2}{q^2} \right) \qquad\qquad \text{(MDM)}\,, \\
& \simeq & \frac{1}{5} + \frac{4 \alpha^2 m_e^2 }{5 q^2} \qquad \qquad m_\chi\gg m_e,~v_\text{rel}\simeq \alpha \, ,
\end{eqnarray}
which is a combination of $F_\textrm{DM}=1$ and $F_\textrm{DM}=\alpha m_e/q$. In deriving this form factor, we find an explicit dependence on the relative velocity between the incoming dark matter and the incoming electron in the free $2\to2$ (dark-matter-electron to dark-matter-electron) scattering.  A precise calculation of the crystal form factor defined in~\cite{Essig:2015cda} would need to take this into account.  However, here we approximate $v_\text{rel} \simeq \alpha$ and calculate the direct-detection bounds and direct-detection projections using 
\begin{equation}
\overline\sigma_e^{\text{MDM}}\simeq 5 \left( \sigma^{-1}_{F_\textrm{DM}=1} + \frac{4\alpha^2 m_e^2}{q^2} \sigma^{-1}_{F_\textrm{DM}=\alpha m_e/q} \right)^{-1}\,.
\end{equation}
Finally, to convert nuclear recoil cross section sensitivities to $\overline\sigma_e$, we follow~\cite{Essig:2019kfe}.  

We show in Fig.~\ref{fig:summaryHP} and in Fig.~\ref{fig:summaryEDM} the sensitivity (when available in the literature) for a few future planned direct-detection and fixed-target experiments or proposals: a silicon detector with a 30-kg-year exposure and single-electron threshold (using, for example, Skipper-CCDs~\cite{Tiffenberg:2017aac,Essig:2015cda}), a superfluid helium detector with a 1~kg-year exposure and 10~eV phonon energy threshold~\cite{Hertel:2018aal}, 
an electron-beam fixed-target experiment searching for missing momentum (LDMX, from Fig.~5 in~\cite{Akesson:2018vlm}), and an electron-positron collider searching for missing energy (Belle-II)~\cite{Essig:2013vha} (the latter two do not have sensitivity to dipole moment dark matter in the range of parameters shown in the plot~\cite{Chu:2018qrm}).  
In Fig.~\ref{fig:summaryHP}, we also show in gray the bound from $N_{\rm eff}$ (also seen in Fig.~\ref{fig:stellarConsHP}) as well as current laboratory bounds from direct-detection and accelerator-based probes, including XENON10/100/1T, DarkSide-50, DAMIC-SNOLAB, SENSEI, SuperCDMS, E137, LSND, and BaBar~\cite{Essig:2012yx,Essig:2013vha,Batell:2014mga,Crisler:2018gci,Agnese:2018col,Abramoff:2019dfb,Agnes:2018oej,Essig:2017kqs,Abdelhameed:2019hmk,Aprile:2019xxb,Aguilar-Arevalo:2019wdi}. In Fig.~\ref{fig:summaryEDM}, we show in gray the bound from $N_{\rm eff}$ (also seen in Fig.~\ref{fig:stellarConsEDM}) as well as the direct-detection bounds from~\cite{Essig:2017kqs,Aguilar-Arevalo:2019wdi}.  At low couplings, the limit reaching to $\sim$100~MeV is from supernova 1987A (from~\cite{Chang:2018rso} for the dark photon portal, and from~\cite{Chu:2019rok} for dark matter coupled to an electric or magnetic dipole moment); we find that the couplings that are probed by the supernova bound lie between the freeze-in and the freeze-out line.  
The CMB (dotted gray line) sets a strong constraint for Dirac fermion dark matter, but is easily avoided, for example, for scalar dark matter~\cite{Madhavacheril:2013cna,Liu:2016cnk}. 
The freeze-out line in Fig.~\ref{fig:summaryHP} is almost independent of the dark photon mass as long as the dark photon mass is sufficiently far away from $2m_\chi$, so we just present the line for the benchmark case $m^\prime=3m_\chi$ and $\alpha_D=0.5$ (although see~\cite{Feng:2017drg}). 
We repeat the freeze-in lines from Figs.~\ref{fig:FIfermion} and \ref{fig:thermalDipoleM} as well as the stellar cooling and other bounds from Figs.~\ref{fig:stellarConsHP} and \ref{fig:stellarConsEDM}.  
As expected, the freeze-in parameters are typically too small to be probed by laboratory searches in the near future.  However, interestingly, we see that the freeze-in targets for dark matter interacting with an electric or magnetic dipole moment \textit{can} be probed for low reheating temperatures with upcoming direct-detection experiments. 

Since the dark matter models with an electric or magnetic dipole moment are dimension~5 operators, one can ask how these are UV completed. As discussed in Sec.~\ref{sec:models}, one simple possibility is to imagine charged scalars and fermions of a common mass $M$ generating the dipole moment operators.  In Fig.~\ref{fig:summaryEDM}, we show the resulting upper bound on the cross section $\overline\sigma_e$ on this simple UV completion, derived from the upper bound on $d_\chi$ or $\mu_\chi$ from Eq.~(\ref{eq:dipole-upper-bound}): the black (gray) solid line corresponds to cross sections above which the corrections to the dark matter mass is larger than the tree-level dark matter mass for $M$ at the scale of 100~GeV (1~TeV).  Of course, different UV completions may allow for higher cross sections.  We do not consider this further in this paper.  

\section{Conclusions}\label{sec:conclusions}
In this work, we discussed the freeze-in production of dark matter in the keV-to-GeV mass range as well as the constraints from stellar cooling.  We considered two distinct scenarios: fermionic and bosonic dark matter that is coupled to the SM  through kinetic mixing between the photon and a dark photon, as well as fermionic dark matter interacting with SM photons through an electric or magnetic dipole moment.

When the dark matter interactions with SM particles are small, the dark sector is not in thermal equilibrium with the SM in the early Universe, and dark matter production can occur through freeze-in from fermion-antifermion annihilation and from the decay of plasmons. The latter dominates for sub-MeV masses and pushes the couplings needed to obtain the observed relic abundance from freeze-in to very small values. For the dark photon portal models, the production in the early Universe is infrared dominated. In contrast, the dark matter models with a dipole moment are described by dimension-five operators, so that the freeze-in production occurs at all temperatures. Consequently, the freeze-in parameters depend on the reheating temperature.  %Moreover, the production from longitudinal plasmon modes becomes more important at higher reheat temperatures.
We also checked that the dark matter does not thermalize with the SM thermal bath, such that the bounds from BBN and $N_\text{eff}$, which usually constrain dark matter with masses in the sub-MeV-range, are avoided. 

In addition to deriving the freeze-in production, we also calculated the stellar cooling constraints, and find that the strongest limits are from the non-delay of helium ignition at the red giant tip.  This bound excludes freeze-in production of dark photon portal dark matter with masses below, for example, 20~keV for a dark gauge coupling of $\alpha_D=0.5$. For dark matter with an electric or magnetic dipole moment with $T_\text{RH}>1$~GeV, dark matter masses below 4~keV are disfavored from stellar cooling constraints. 

Finally, we discussed the potential to probe these models in laboratory experiments. In the case of dark matter coupled via a dark photon, some part of the parameter space that can be probed by future experiments is already ruled out by the red giant constraint. Towards larger dark matter masses, the freeze-in lines are too low to be probed in the foreseeable future, but present potential targets for future, very ambitious, experiments.  However, for dark matter with an electric or magnetic dipole moment, and for dark matter masses above the reheating temperature, the freeze-in production in the early Universe is suppressed; relatively large couplings are required to then obtain the correct relic abundance, so that these scenarios can be partially probed with upcoming direct-detection experiments.  

\section*{Acknowledgements}
We thank Amitayus Banik, Herbi Dreiner, Felix Kahlhoefer, Johannes Herms, Saurabh Nangia, Javier Redondo, and especially Josef Pradler for helpful discussions and/or correspondence. 
We also thank Shiuli Chatterjee, Xiaoyong Chu, Simon Knapen, Eric Kuflik, Jui-Lin Kuo, Ranjan Laha, Tongyan Lin, Josef Pradler, and Lukas Semmelrock
for correspondence and/or discussions that prompted us to correct some errors in our calculations in previous versions of this paper.  
R.E.~and J.H.C.'s~work in this paper is supported by DoE Grant DE-SC0017938. R.E.~also acknowledges support
from the US-Israel Binational Science Foundation under Grant No.~2016153, from the Heising-Simons Foundation
under Grant No.~79921, from a subaward for the DOE Grant No.~DE-SC0018952, and from Simons Investigator
Award 623940.
AR is supported by the Cusanuswerk. 

\appendix
\section{Properties Of Photons In A Thermal Plasma}\label{app:plasmaProp}
In a plasma, electrons can move freely and thus affect the propagation of electromagnetic waves. They become a combination of coherent vibrations of not only the electromagnetic field, but also the electron density. Quantization leads to a spin-1 field with one longitudinal and two transverse polarization modes.  We review the material needed to derive our results in this paper, basing our discussion on~\cite{Braaten:1993jw}. 

The effectively massive photon modes are caused by a modified dispersion relation for the photon in a plasma. For a photon in vacuum, the relation between its frequency $\omega$ and wave vector $\vec{k}$ is simply given by $\omega^2=k^2$. For plasmons, this relation is subject to modifications depending on the electron density $n_e$ and temperature $T$. The modified dispersion relations give rise to a non-zero phase-space $\omega^2-k^2$ allowing for decays to massive particles.
Note that in principle free protons and nuclei could also contribute to the plasma effect. However, they are much heavier than the electrons and thus more inert, so their contribution turns out to be negligible.

A characteristic quantity of a plasma is its plasma frequency
\be\label{eq:omegaP}
\omega_p^2=\frac{4\alpha}{\pi}\int_0^\infty \dd p \frac{p^2}{E}\lb 1-\frac{v^2}{3}\rb \lb n_e(E)+\bar{n}_e(E)\rb\ .
\ee
It is in general a function of the temperature $T$, as the electron (positron) density follows the Fermi distribution $n_{e/\bar{e}}=[e^{(E\mp \mu)/T}+1]^{-1}$ with the chemical potential $\mu$.
For the explicit computation it is helpful to replace $v=\frac{p}{E}$.
Defining 
\be
\omega_1^2=\frac{4\alpha}{\pi}\int_0^\infty \dd p \frac{p^2}{E}\lb \frac{5}{3}v^2-v^4\rb \lb n_e(E)+\bar{n}_e(E)\rb\ ,
\ee
allows the definition of the quantity $v_\star=\omega_1/\omega_p$, which intuitively is the typical electron velocity. 
With these ingredients, the general dispersion relations valid at all temperatures and densities up to first order in the electromagnetic fine structure constant $\alpha$ are given by~\cite{Braaten:1993jw} 
\be
\omega_T^2&=&k^2 +\omega_p^2\frac{3\omega_T^2}{2\vst^2k^2}\left(1-\frac{\omega_T^2 -\vst^2k^2}{\omega_T^2}\frac{\omega_T}{2\vst k}\ln\left(\frac{\omega_T +\vst k}{\omega_T -\vst k}\right)\right),\quad 0\leq k <\infty \label{eq:dispersionT} \\
\omega_L^2&=&\op^2\frac{3\oml^2}{\vst^2 k^2}\left(\frac{\oml}{2\vst k}\ln\left(\frac{\oml +\vst k}{\oml -\vst k}\right)-1\right), \quad 0\leq k < k_\text{max}\label{eq:dispersionL}\ .
\ee
The transverse mode satisfies $\omega_T>k$ for all values of $k$. In contrast, the dispersion relation for the longitudinal mode can cross the light cone if $k$ becomes larger than $\omega_L$. This prevents the longitudinal plasmon from propagating and constrains the longitudinal wave vector to a maximal value
\be
k_\text{max}&=&\frac{4\alpha}{\pi}\int_0^\infty \dd p \frac{p^2}{E}\lb \frac{1}{v}\ln\lb\frac{1+v}{1-v}\rb -1\rb \lb n_e(E)+\bar{n}_e(E)\rb\\
&=&\left[\frac{3}{\vst^2}\lb \frac{1}{2\vst}\ln\lb\frac{1+\vst}{1-\vst}\rb-1\rb\right]^{1/2}\op\ .
\ee

The renormalization of the propagator determines the propagation of plasmons. However, when interactions are considered it is useful to change from the mass to the interaction basis. The coupling to the electromagnetic current then gets renormalized. 
\begin{align}
Z_T(k)&=\frac{2\omt^2 (\omt^2-\vst^2k^2)}{3\op^2\omt^2+(\omt^2+k^2)(\omt^2-\vst^2k^2)-2\omt^2(\omt^2-k^2)}\ , \\
\quad Z_L(k)&=\frac{2(\oml^2-\vst^2k^2)}{3\op^2-(\oml^2-\vst^2k^2)}\frac{\oml^2}{\oml^2-k^2}\,,
\end{align}
such that the dressed polarization vectors are~\cite{Chu:2019rok}
\begin{align}
\tilde{\epsilon}^\mu_T=\sqrt{Z_T}\epsilon^\mu_T\ ,\qquad \tilde{\epsilon}^\mu_L=\sqrt{Z_L}\epsilon^\mu_L\,.
\end{align}

We now want to discuss specific limits that are helpful for our numerical implementation of the calculations. 
In general, as $k\rightarrow 0$ the dispersion relations $\omega_{t/l}$ approach the plasma frequency. For large wave numbers $k\gg T$ and small electron density, the situation of the vacuum is restored, $\omega_T\rightarrow k$ and the longitudinal mode disappears.

In the \textbf{relativistic limit}, $T\gg m_e$ or $\mu\gg m_e$, Eqs. \eqref{eq:dispersionT} and \eqref{eq:dispersionL} simplify as $\vst=1$ and $k_\text{max}\rightarrow\infty$.
The plasma frequency reduces to 
\be
\omega_{p, \text{rel.}}^{2}=\frac{4\alpha}{3\pi}\lb \mu^2+\frac{\pi^2 T^2}{3}\rb\ .
\ee

In the \textbf{degenerate limit}, $T\ll \mu-m_e$, the plasma frequency can be expressed in terms of the Fermi momentum $p_F$
\begin{align}
\omega_{p, \text{deg.}}^{2}=\frac{4\alpha}{3\pi}p_F^2v_F\ ; \quad p_F=\lb 3\pi^2 n_e\rb^{1/3}\ .
\end{align}
In the dispersion relations, $\vst$ can be replaced by the Fermi velocity $v_F=\frac{p_F}{E_F}$ with the Fermi energy $E_F=\sqrt{p_F^2+m_e^2}$.

In the \textbf{classical limit}, the electrons are non-relativistic and non-degenerate, $T\ll m_e-\mu$. The plasma frequency is given by
\be
\omega_{p, \text{cl.}}^{2}=\frac{4\pi\alpha n_e}{m_e}\lb 1-\frac{5}{2}\frac{T}{m_e}\rb\,,
\ee
and the dispersion relations reduce to
\be
\omt^2&=&k^2+\omega_p^2\lb 1+\frac{k^2}{\omt^2}\frac{T}{m_e}\rb\ ,\quad 0\leq k <\infty\\
\oml^2&=&\op^2\lb 1+ 3\frac{k^2}{\oml^2}\frac{T}{m_e}\rb\ ,\quad 0\leq k <\op\sqrt{1+3T/m_e} \ .
\ee

Most contributions to the freeze-in for the dark photon portal comes from late stages of the thermal history of the early Universe. For dark matter masses below the electron mass, the classical limit is important as production occurs partly when the electrons are non-relativistic. At temperatures of tens of keV, the lepton asymmetry becomes important, such that the sum of the electron and positron number densities is given by
\begin{align}
n_e^\text{non-rel.}=4\left(\frac{m_e T}{2\pi}\right)^{3/2}\exp\left(-\frac{m_e}{T}\right)+\eta_B n_\gamma\ .
\end{align}
In the last term, $\eta_B\approx 6\times 10^{-10}$ is the baryon to photon ratio and $n_\gamma=2\zeta_3 T^3/\pi^2$ is the photon number density with the Riemann zeta function value $\zeta_3\approx 1.2$. Since the baryon number density seems to coincide with the number density of electrons, the last term accounts for the asymmetry.

\section{Inclusion of $A' - Z$ Mixing In Freeze-In Calculations}\label{app:Z-boson}
The mixing term between the dark photon and the hypercharge gauge boson in the Lagrangian reads
\begin{equation}
\mathcal{L} \supset -\frac{\epsilon}{2\cos\theta_W}F^\prime_{\mu\nu}B^{\mu\nu}\ .
\end{equation}
After the electroweak symmetry breaking, this term can be written with gauge boson mass eigenstates, 
\begin{equation}
\mathcal{L} \supset-\frac{\epsilon}{2}F^\prime_{\mu\nu}F^{\mu\nu}-\frac{\epsilon \tan\theta_W}{2}F^\prime_{\mu\nu}Z^{\mu\nu}\ ,
\end{equation}
where $Z_{\mu\nu}$ is the field strength of the $Z$ boson. The second term is negligible at low energies, but can be relevant for energies larger than the GeV scale. In this work, we mainly focus on the sub-GeV scale, so the contribution from $Z$-mixing is less than $\mathcal{O}(10\%)$. However, we include the contribution from $Z$-mixing in our calculations, and briefly summarize the relevant formula in this Appendix. For the $Z$-mixing contribution, we ignore plasma effects because the effects do not open a new production channel, and the correction is not significant. Also, we do not include $Z$-mixing for the stellar bounds, as the temperature of the stellar objects are very small compared to the $Z$-boson mass.

\subsection{$Z$-Boson Decay}
The last term in Eq.~(\ref{eq:ratedp}) describes the contribution from the $Z$-boson decay to a dark matter pair, which dominates for $10~\text{GeV} \lesssim m_\text{DM} < m_Z/2$. The term for the case of fermionic dark matter $\chi$ can be written as
\begin{equation}
n_{Z}\langle\Gamma\rangle_{Z\rightarrow \chi \bar\chi} = \frac{g_Z m_Z^2 T}{2 \pi^2} \Gamma_{Z \rightarrow \chi \bar\chi} K_1\left(\frac{m_Z}{T} \right)\ ,
\end{equation}
where $g_Z=3$ is the degrees of freedom of the $Z$ boson, and 
\begin{equation}
\Gamma_{Z \rightarrow \chi \bar\chi} = \frac{1}{3}\alpha_D \epsilon^2 \tan\theta_W^2 m_Z \left( 1+2 \frac{m_\chi^2}{m_Z^2} \right) \sqrt{1-4 \frac{m_\chi^2}{m_Z^2} }\ .
\end{equation}

\subsection{Annihilation through the $Z$ boson}
In Eq.~(\ref{eq:intMatrixScattFermion}), we only show the amplitude for production through the photon. Here, we show the full amplitude with the $Z$-boson: 
\begin{align}
\int\dd\Omega\left|\mathcal{M}\right|^2=&\frac{16 \pi}{3}\frac{(\epsilon e q_f g_D)^2 (s+2m_f^2)(s+2m_\chi^2)}{(s-m^{\prime 2})^2+m^{\prime 2}\Gamma_{A^\prime}^2}\nonumber \\
&+\frac{4\pi}{3}\frac{(\epsilon \tan \theta_W g_Z g_D)^2 s^2 (s+2m_\chi^2)\left[ \left(C_{V}^{f}\right)^2 \left(s+2m_f^2 \right) + \left(C_{A}^{f}\right)^2 \left(s - 4 m_f^2 \right) \right]}{\left((s-m^{\prime 2})^2+m^{\prime 2}\Gamma_{A^\prime}^2\right)\left((s-m_Z^{2})^2+m_Z^{2}\Gamma_{Z}^2\right)} \nonumber\\
&+\frac{16\pi}{3}\frac{(\epsilon^2 \tan \theta_W e q_f g_Z C_V^f g_D^2) s (s+2m_\chi^2)(s+2m_f^2)(s-m_Z^2)}{\left((s-m^{\prime 2})^2+m^{\prime 2}\Gamma_{A^\prime}^2\right)\left((s-m_Z^{2})^2+m_Z^{2}\Gamma_{Z}^2\right)}\ ,
\end{align}
where $\theta_W$ is the weak mixing angle, $g_Z=\frac{e}{\cos \theta_W \sin \theta_W}$, $C_V^f = T_f^3-2q_f \sin^2 \theta$, $C_A^f = T_f^3$, and $\Gamma_Z \simeq 2.5 \text{GeV}$ is the decay width of the $Z$-boson.

\section{Bremsstrahlung Processes}\label{app:Brems}
In this appendix, we calculate the energy loss rate through the bremsstrahlung processes discussed in Sec.~\ref{subsubsec:brem} ($e(P_1) + p (P_2) \rightarrow e(P_3) + e(P_4) + \chi (P_\chi) + \bar\chi (P_{\bar\chi}$)) asssuming the soft radiation approximation (SRA). The expression for the energy loss rate is
\begin{equation}\label{eq:bremratedef}
\frac{\dd L_\chi^\textrm{Brems}}{\dd V} = \int \dd \Pi_1 \dd \Pi_2 \dd \Pi_3 \dd \Pi_4 \dd \Pi_\chi \dd \Pi_{\bar\chi} (2 \pi)^4 \delta^4(P_1+P_2-P_3-P_4-K) \, \omega \, |\mathcal{M}|_\text{Brems}^2,
\end{equation}
where $\dd \Pi$ is defined in Eq.~\eqref{eq:dPi} and $K=(\omega,\vec{k})$ is the 4-momentum of the radiated off-shell photon. By using SRA, we get
\begin{eqnarray}
\delta^4(P_1+P_2-P_3-P_4-K) &=& \delta^4(P_1+P_2-P_3-P_4) e^{-\omega/T} \quad\text{and} \label{eq:c2}\\
|\mathcal{M}|_\text{Brems}^2 &=& 4\pi \alpha |\mathcal{M}|_{ep}^2 \frac{L_\mu L_\nu}{s_\chi^2} \mathcal{T}^{\mu \nu},
\end{eqnarray}
where
\begin{eqnarray}
L_\mu &=& \frac{P_1}{P \cdot K} - \frac{P_3}{P_3 \cdot K}, \\
s_\chi &=& \omega^2 - k^2,\\
\mathcal{T}^{\mu \nu} &=& \text{Tr}\left[(\slashed{p}_\chi + m_\chi) \Gamma^\mu (\slashed{p}_{\bar\chi}-m_\chi) \Gamma^\nu \right], \label{eq:c6}
\end{eqnarray}
$T$ is the temperature, and $\Gamma^\mu$ is the vertex factor of the interaction between $\chi$ and the photon. 
Inserting Eqns.~(\ref{eq:c2}) to (\ref{eq:c6}) and 
\begin{equation}
1 = \int \frac{\dd^4 k}{(2 \pi)^4} (2 \pi)^4 \delta^4(K-P_\chi-P_{\bar\chi})
\end{equation}
into Eq.~\eqref{eq:bremratedef} yields
\begin{eqnarray}\label{eq:Lbrem2}
\frac{\dd L_\chi^\textrm{Brems}}{\dd V} &=&  \int \dd \Pi_1 \dd \Pi_2 \dd \Pi_3 \dd \Pi_4 (2 \pi)^4 \delta^4(P_1+P_2-P_3-P_4) |\mathcal{M}|_{ep}^2 \nonumber\\
&&\times \int \frac{\dd^4 k}{(2 \pi)^4} 4\pi \alpha \, \omega \, e^{-\omega/T} \frac{L_\mu L_\nu}{s_\chi^2} \nonumber\\
&&\times \int \dd \Pi_\chi \dd \Pi_{\bar\chi} (2 \pi)^4 \delta^4(K-P_\chi-P_{\bar\chi}) \mathcal{T}^{\mu \nu}.
\end{eqnarray}
The last line in Eq.~\eqref{eq:Lbrem2} is calculated in \cite{Chu:2019rok}, which is
\begin{equation}\label{eq:Imunu}
\int \dd \Pi_\chi \dd \Pi_{\bar\chi} (2 \pi)^4 \delta^4(K-P_\chi-P_{\bar\chi}) \mathcal{T}^{\mu \nu} = \frac{1}{8\pi} \sqrt{1-\frac{4m_\chi^2}{s_\chi}}f_\text{DM}(s_\chi)\left ( -g^{\mu\nu}+\frac{K^\mu K^\nu}{s_\chi} \right ),
\end{equation}
with $f_\text{DM}(s)$ is shown in Eqs.~\eqref{eq:fDMDP}-\eqref{eq:fDMMDM}. Using Eq.~\eqref{eq:Imunu},
\begin{eqnarray}
\frac{1}{4\pi} \int \dd \Omega_{k} L_\mu L^\mu &=& \frac{3\omega^2-k^2}{3\omega^4} \frac{|\vec{p}_1-\vec{p}_3|^2}{m_e^2}, \quad \text{and} \\
\int \frac{\dd^4 k}{(2 \pi)^4} &=& \frac{1}{2\pi}\int \dd \Pi_k \dd s_\chi,
\end{eqnarray}
we get Eq.~\eqref{eq:bremrate}. Note that we ignore factors of $(1-f)$ in $\dd\Pi$ for all particles except electrons (i.e., for protons, photons, and the dark matter) in the calculations.

\bibliography{ref}
\bibliographystyle{jhep}

\end{document}